\documentclass[a4paper,11pt]{article}

\usepackage{amssymb,latexsym} 
\usepackage{graphicx}

\usepackage{amsmath,amssymb,amsthm}
\usepackage{color}
\usepackage{verbatim}
\usepackage{vmargin}
\usepackage{url}
\usepackage[english]{babel}
\usepackage[latin1]{inputenc}
\usepackage[T1]{fontenc}
\usepackage{enumitem}
\usepackage{array}
\usepackage{fancyhdr}
\usepackage{lmodern}

\begin{document}

\thispagestyle{empty}
\begin{center}
\textbf{ \large{Bifurcation analysis of the Yamada model for a pulsing semiconductor laser with saturable absorber and delayed optical feedback}}
\vspace{5mm}

Soizic Terrien \footnote{The Dodd-Walls Centre for Photonic and Quantum Technologies, Department of Mathematics, The University of Auckland, New Zealand. corresponding author: \url{s.terrien@auckland.ac.nz}}, Bernd Krauskopf \footnotemark[1], Neil G.R. Broderick \footnote{The Dodd-Walls Centre for Photonic and Quantum Technologies, Department of Physics, The University of Auckland, New Zealand.}

\end{center}

\subsubsection*{Abstract}
\begin{small}
Semiconductor lasers exhibit a wealth of dynamics, from emission of a constant beam of light, to periodic oscillations and excitability.
Self-pulsing regimes, where the laser periodically releases a short pulse of light, are particularly interesting for many applications, from material science to telecommunications. Self-pulsing regimes need to produce pulses very regularly and, as such, they are also known to be particularly sensitive to perturbations, such as noise or light injection.

\noindent
We investigate the effect of delayed optical feedback on the dynamics of a self-pulsing semiconductor laser with saturable absorber (SLSA). More precisely, we consider the Yamada model with delay -- a system of three delay-differential equations (DDEs) for two slow and one fast variable -- which has been shown to reproduce accurately self-pulsing features as observed in SLSA experimentally. 
This model is also of broader interest because it is quite closely related to mathematical models of other self-pulsing systems, such as excitable spiking neurons. 

We perform a numerical bifurcation analysis of the Yamada model with delay, where we consider both the feedback delay, the feedback strength and the strength of pumping as bifurcation parameters. We find a rapidly increasing complexity of the system dynamics when the feedback delay is increased from zero. In particular, there are new feedback-induced dynamics: stable quasi-periodic oscillations on tori, as well as a large degree of multistability, with up to five pulse-like stable periodic solutions with different amplitudes and repetition rates.
An attractor map in the plane of perturbations on the gain and intensity reveals a Cantor set-like, intermingled structure of the different basins of attraction. This suggests that, in practice, the multistable laser is extremely sensitive to small perturbations. 
\end{small}

\subsubsection*{Key words}
\begin{small}
Semiconductor lasers with saturable absorber, Delay-differential equations, Bifurcation analysis, Multistability. 
\end{small}

\section{Introduction}
\label{sec:intro}

Semiconductor lasers are very efficient sources of light, widely used for many different applications, such as telecommunications \cite{Ohtsubo_in_Kane_05}, cryptography \cite{Ludge_book}, optical data storage \cite{Mossberg_82} and eye surgery \cite{Webb_04}. This popularity is easily explained by the features of semiconductor lasers: they are small, cheap, and efficient \cite{Agrawal}.
Semiconductor lasers have thus attracted considerable attention in the last decades, and their dynamics have been studied widely, both from theoretical and experimental points of view; see for example \cite{Kane_book, Krauskopf_Lenstra_book, Ludge_book} as entry points to the extensive literature on laser dynamics. Apart from stable emission of a continuous beam of light, semiconductor lasers have been shown to produce a wealth of other dynamics. This includes excitability \cite{Dubbeldam_excitability_99, Giudici_excitability_97, Tredicce_excitability_expe, Krauskopf_OpticsComm_2003}, which corresponds to a all or none, well-calibrated response to a perturbation, periodic oscillations \cite{Abraham_overview85, Dubbeldam_99, Erneux_JOSA_88}, quasiperiodic modulation of the light amplitude \cite{Pieroux_PRL_01, Winfull_APL_86}, as well as chaotic dynamics \cite{Mork_92,Shore_87}. Pulse-like periodic solutions, where the laser periodically produces a very short, high-amplitude pulse of light, are particularly interesting from an application point of view. 
However, trains of pulses may be very sensitive to perturbations, such as noise \cite{Georgiou_92}, optical feedback \cite{Krauskopf_book_chapter}, or injection of the light produced by another laser \cite{Wieczorek_99}.  A small amount of internal or external noise can induce timing-jitter of the pulses  \cite{Jaurigue_15, Otto_NJP_2012,vanTartwijk_96}: pulses are then produced with a non-constant repetition rate, which is detrimental to many applications.

\begin{figure}[t!]
\begin{center}
\includegraphics[width=0.7\textwidth]{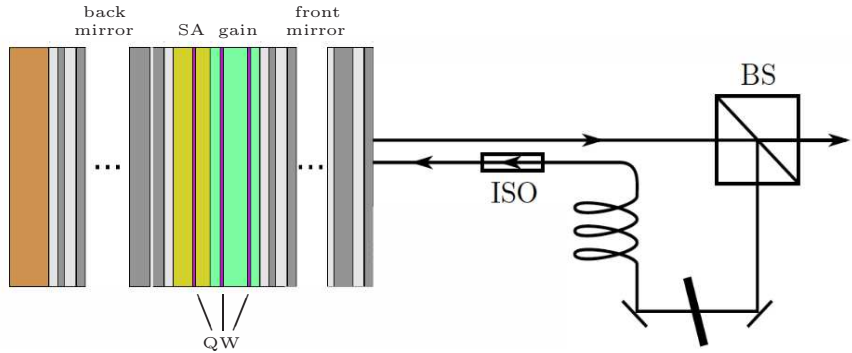}
\caption{Sketch of a semiconductor laser with saturable absorber (from \cite{Barbay_OL_2011}), where a fraction $\kappa$ of the output light is fed back after a delay $\tau$ determined by the length of the delay loop. SA: saturable absorber, BS: beam splitter, ISO: optical isolator, QW: quantum wells.}
\label{fig:VCSEL}
\end{center}
\end{figure}

We focus here on a micropillar semiconductor laser with saturable absorber (SLSA) subject to delayed optical feedback. Figure \ref{fig:VCSEL} is a sketch of the device: the laser consists of an optical resonant cavity of a few microns bounded by two parallel, high-reflectivity mirror stacks.  The top facet where the light is emitted is of such small diameter (3 $\mu m$) that lasing occurs only in the fundamental mode. The microcavity embeds a gain medium, consisting of two quantum wells \cite{Elsass_Eur_Phys_2010}. The energy provided to the system through optical pumping is stored in that gain section, which can then act as a light amplifier. The key feature of the device is that it includes a saturable absorber section in the cavity, consisting of a single quantum well  \cite{Elsass_Eur_Phys_2010}. In such an absorber, the optical transmission increases when the light intensity is larger than a given threshold.
Without feedback, this device has been shown, both experimentally and theoretically, to exhibit excitability \cite{Barbay_OL_2011, Dubbeldam_excitability_99, Selmi_PRL_14} and self-pulsations \cite{Elsass_Eur_Phys_2010,Erneux_JOSA_88} arising through an homoclinic bifurcation \cite{Dubbeldam_99}. 

Different physical mechanisms can induce self-pulsations in lasers. Mode-locking relies on a phase synchronisation of the different longitudinal modes of the laser cavity, leading to the periodic emission of extremely short pulses of light with high repetition rate  \cite{Rafailov_07}. 
On the other hand, a laser as discussed here is essentially a single longitudinal mode system, so that mode-locking cannot take place. For this type of device, self-pulsations occur through a passive Q-switching mechanism \cite{Erneux_JOSA_88}, which results from variable losses in the cavity and leads to the emission of a train of high-intensity light pulses. In the case of passive Q-switching, variable losses are induced by the saturable absorber. The physical mechanism can be explained schematically as follows. At the first stage, the intracavity intensity is low, and the losses due to the saturable absorber are consequently high. The energy provided by pumping the gain medium is then almost entirely stored in the gain section. When the gain becomes sufficiently high to overcome the losses, the intracavity intensity starts to grow, and eventually saturates the absorber, which results in a sudden drop of the cavity losses. As the energy stored in the gain section is high, the intensity can grow quickly, until all the stored energy is released as a pulse of light. In the process, the intracavity intensity drops back to a low value, and the same process can repeat. However, the next pulse can only be triggered after a given amount of time, called the refractory period, during which the gain section recovers; see, for example, \cite{Selmi_PRL_14}. Because the electric field in the cavity evolves on a shorter time-scale than the carrier population in gain and absorber media, one obtains a train of short light pulses separated by long periods with almost-zero intensity.

The Yamada model \cite{Yamada_93}, a system of three ordinary differential equations (ODEs) for the gain $G$, the absorption $Q$ and the intensity $I$, is a well known model for Q-switched single-mode lasers. This model has been extensively studied; in particular, a complete bifurcation analysis has been performed in \cite{Dubbeldam_99}, where all the possible different dynamics have been determined. It has been shown recently to describe accurately, both qualitatively and quantitatively, a range of dynamics of the micropillar laser sketched in figure \ref{fig:VCSEL} in the case without feedback: this includes self-pulsations, excitability and temporal summation  \cite{Barbay_OL_2011, Selmi_PRL_14, Selmi_OL_15}.

We consider in this study the effect of delayed optical feedback on the SLSA dynamics. Since feedback can either occur naturally in laser systems through unintentional reflections, or be introduced in an attempt of control in many physical systems \cite{Pyragas_92}, the effects of feedback on a laser's dynamics have attracted considerable attention in the last decades; see, for example, \cite{Kane_book} as an entry point to the extensive literature. Different kinds of feedback can be considered, including filtered or phase-conjugate optical feedback \cite{Krauskopf_PRE_98}. However, we consider here the simplest case of conventional optical feedback from a regular mirror, where a fraction $\kappa$ of the output light is reinjected into the laser cavity after a given delay $\tau$; see the sketch in Figure \ref{fig:VCSEL}. More precisely, a beam splitter separates the output light into two beams: the first one constitutes the output of the SLSA, and the other part is reinjected into the cavity, after a delay time $\tau$ that is directly related to the length of the feedback loop. An optical isolator prevents unwanted backward reflections. In Figure \ref{fig:VCSEL}, the feedback loop is sketched with an optical fiber for sake of clarity; however, the delay loop is realised in the experiment through propagation in free space \cite{Felix_master_thesis}.
Our focus is the effect of delayed optical feedback on the excitable regime, which has been shown to occur, in a SLSA, just below the laser threshold at which the laser starts to emit light and then produces self-pulsations \cite{Dubbeldam_99}. The motivation for adding a feedback loop into the excitable SLSA relies on the following simple idea: when an excitable pulse in reinjected after a delay $\tau$, it will trigger a new pulse if the reinjection strength $\kappa$ is large enough. As this process repeats, a regular pulse train will be produced, where the repetition rate is directly related to the delay time $\tau$, and the pulse amplitude and shape are well-calibrated. By adding a delayed optical feedback loop, we aim here at achieving a better control of the self-pulsing solutions: this includes the reduction of timing-jitter, as well as the control of amplitude and repetition rate.

We consider here the Yamada model with an additional delayed term to describe the delayed optical feedback, as already introduced in \cite{Krauskopf_book_chapter}. In dimensionless form, it can be written as the system of three delay-differential equations for the gain $G$, the absorption $Q$ and the intensity $I$:
\begin{equation}
\begin{split}
\dot G &= \gamma \left[A-G-GI\right] ,\\
\dot Q &= \gamma\left[B-Q-aQI\right], \\
\dot I &= \left[G-Q-1\right]I + \kappa I(t-\tau).
\end{split}
\label{eq:Yamada_feedback}
\end{equation} 
Here, $A$ is the pump parameter of the gain, $B$ describes the nonsaturable losses, $a$ is the saturation parameter, and $\gamma$ is the recombination rate of carriers in the gain and absorber medium (which are assumed here to be the same). In the feedback term of the intensity equation, $\kappa$ is the feedback strength, and $\tau$ is the feedback delay. For physical reasons, all these parameters are positive. The time variables are rescaled to the cavity photon life time, which has been estimated to be of the order of a few picosecond for the device we consider here \cite{Barbay_OL_2011}.

Many models for semiconductor lasers with delay include the complex electric field $E(t)$ and not only the intensity as discussed here. One speaks of Lang-Kobayashi-type rate equation models \cite{Jaurigue_15,Lang_Kobayashi,Otto_NJP_2012}. We consider here a situation in which the SLSA is either off, in an excitable regime, or produces pulses. In between such very short light pulses, the electric field inside the laser is thus almost zero for long periods of time, its only contribution being the spontaneous emission noise (which is not considered in (\ref{eq:Yamada_feedback})). Such a configuration is conceptually very different from optical feedback introduced into a laser that operates in or near a continuous-wave regime.  More precisely, the response of the excitable SLSA to a sufficiently large external perturbation is a high-amplitude pulse of approximately 200 $ps$ duration \cite{Selmi_PRL_14}. When an additional feedback loop is considered, the feedback delay is in practice considerably larger than the pulse duration (typically between 4 $ns$ and 60 $ns$). When the excitable pulse is reinjected, the intracavity field has thus come back to an almost-zero value and has low coherence. Consequently, its phase is actually not properly defined. Indeed, it has been shown numerically that the feedback phase is not relevant in the limit of low to medium pumping \cite{vanTartwijk_96}. In this specific configuration, one can thus reasonably consider that the electric field $E(t)$ at the instant $t$ does not interfere with the reinjected, delayed field $E(t-\tau)$. Conceptually, this configuration is very similar to the case, without feedback, when the excitable SLSA is subject to an external pulse perturbation. It has been shown experimentally that then the system dynamics only depends on the amplitude of the perturbation: the response of the device to a perturbation is qualitatively and quantitatively the same whether this perturbation is to the gain $G$ or to the intensity $I$ \cite{Selmi_OL_15}. In particular, the feedback is incoherent and linear in the intensity \cite{Pieroux_PRA_94,Solorio_incoherent_FB_02}.
The Yamada model with an additional feedback term in the intensity equation thus emerges as a natural first minimal model for the device considered here. Because it is mathematically quite simple, it is possible to study analytically the steady-state solutions \cite{Krauskopf_book_chapter}, as well as to perform an in-depth numerical bifurcation analysis, highlighting a wide range of the different possible, complex dynamics.

Although they are derived from the same physical principles, the Yamada model with feedback and Lang-Kobayashi-type models for semiconductor lasers subject to feedback rely on different assumptions related to the specifics of the considered device; as such they considerably differ from each other \cite{Lang_Kobayashi,Otto_NJP_2012}. In fact, equations (\ref{eq:Yamada_feedback}) can be related directly to the Lang-Kobayashi equations: considering only the intensity dynamics and taking into account the assumption of non-interference between the delayed and instantaneous fields results in a feedback term that is formally different but gives the same results. A more detailed comparison between these Yamada-type and Lang-Kobayashi-type models for pulsing lasers with (self)-feedback is beyond the scope of this article and will be discussed elsewhere. In particular, we remark that any of these laser models with feedback may show the phenomenon of practically independent pulses in the external feedback loop, providing the delay $\tau$ is sufficiently large (larger than considered here). Such solutions are also referred to in the literature as temporal cavity solitons or localised structure; see for example \cite{marconi, Romeira_nature_16}.

The Yamada model should be of more general interest, beyond the specific device sketched in figure \ref{fig:VCSEL}. In fact, it can be seen as an optical analogue to excitable spiking neurons: such systems exhibits related behaviours, including excitability, and are modelled accurately with very similar equations that describe the dynamics of a fast voltage, see for example  \cite{izhikevich_book_07}. In particular, these models do not have a phase variable either. More generally, the Yamada model can be seen as a paradigmatic model for coupled pulsing systems, where the configuration considered here corresponds to the particular case of self-coupling.

In this paper, we adopt a dynamical system point of view and perform a bifurcation analysis of the Yamada model with delayed optical feedback. By considering both the feedback strength $\kappa$ and the feedback delay $\tau$ as bifurcation parameters, we aim at studying extensively the effect of the delayed feedback on the laser's dynamics. In a first bifurcation analysis of this model \cite{Krauskopf_book_chapter}, bifurcation diagrams were computed in the $(\tau,\kappa)$-plane, for two different values of the pump parameter $A$, corresponding to two qualitatively different behaviour of the system without feedback. In the first case, the laser without feedback is in the excitable regime, while it is in the non-excitable off-state in the second case \cite{Dubbeldam_99}. The study in \cite{Krauskopf_book_chapter} has shown interesting new feedback-induced dynamics, including bi-stability of two different pulsing solutions. However, it focused on very low values of the feedback delay of $\tau \leq 60$. We extend here the bifurcation analysis for larger values of the feedback parameters. This is physically relevant, as such large values of the delay are usually considered experimentally \cite{Felix_master_thesis}. Moreover, after considering the bifurcation diagrams in the $(\tau,\kappa)$-plane, we also consider bifurcation diagrams in the $(A,\kappa)$-plane of pump current and feedback strength. Again, this is of particular interest from an experimental point of view: the value of the delay being directly related to the length of feedback loop (see figure \ref{fig:VCSEL}), it is fixed during a given experiment. Conversely, the feedback strength $\kappa$ can be tuned quite easily, and the pump current $A$ is the main control parameter. 

Compared to ordinary differential equations (ODEs), solving delay-differential equations (DDEs) requires to specify as an \textit{initial condition} not only the state at time $t=0$, but its values on a whole time interval $[-\tau,0]$. Consequently, a DDE has an infinite-dimensional phase space (see, for example, the review \cite{Roose2007continuation}), and one has to use specific numerical methods for continuation and bifurcation analysis. We use here the numerical continuation software DDE-Biftool \cite{Engelborghs_02,Engelborghs_01,Sieber_biftool}, which has been used to investigate many different delay systems \cite{Green_02,Keane_15,Vladimirov_04}. As explained above, the intensity $I$, on the one hand, and the gain $G$ and absorption $Q$, on the other hand, evolve on different time-scales. In equations (\ref{eq:Yamada_feedback}), this is reflected by a small value of parameter $\gamma$, which is characteristics of a slow-fast system; see for example \cite{Jones_95}. The analysis performed in this paper is thus also an example of a bifurcation analysis of a slow-fast system with a single delay, which can be of broader interest than only for laser applications.

New important information in this study includes the precise loci of the bifurcations of periodic solutions; such bifurcations can now be continued numerically for DDEs \cite{Sieber_biftool}, thus providing more complete bifurcation diagrams. For both the case of the SLSA in an excitable regime ($A = 6.5$) and in a non-excitable off-state ($A = 5.9$), we perform a bifurcation analysis in the $(\tau, \kappa)$-plane. 
Both cases lead to qualitatively different bifurcation diagrams. Both bifurcation diagrams have in common a self-repeating structure, meaning that they become more and more complex as the delay $\tau$ is increased, which is characteristic feature of delay systems \cite{Yanchuk}. In particular, new feedback-induced dynamics is observed in both configurations for sufficiently large values of $\tau$. 
This includes multistability between up to five pulse-like periodic solutions, with different pulse amplitudes and repetition rates, as well as stable quasiperiodic oscillations on tori. In the case of multistability, we show that the periods of the different periodic solutions (\textit{i.e.} the repetition rate of the pulses) are directly related to the value of the delay time $\tau$. Considering three fixed values of $\tau$, we then perform a bifurcation analysis in the $(A,\kappa)$-plane of the pump parameter and feedback strength. It shows that the increasing complexity and wealth of dynamics observed in the $(\tau, \kappa)$-plane can also be accessed in the $(A,\kappa)$-plane. Finally, we investigate the consequences of feedback-induced multistability. In particular, we show that the basins of attraction associated with the different stable pulsing solutions have an intricate, Cantor set-like structure in the $(\Delta G, \Delta I)$-plane of perturbations on the gain $G$ and the intensity $I$. This highlights the extreme sensitivity of the multistable laser to small perturbations.

The paper is organized as follows. The main results of the first bifurcation study performed in \cite{Krauskopf_book_chapter} are briefly summarized in section \ref{sec:background model} as a starting point. In section \ref{sec:diag tau kappa}, an extended bifurcation analysis of the model in performed in the $(\tau,\kappa)$-plane of feedback delay and feedback strength, for two different values of the pump parameter $A$ corresponding to two qualitatively different configurations. Bifurcation diagrams in the $(A,\kappa)$-plane are studied in section \ref{sec:bif study A kappa}, for three fixed values of $\tau$. Finally, the consequences of multistability are investigated in section \ref{sec:multistability}.  We summarize and point out directions for future work in section \ref{sec:discussion}.

\section{Background on the Yamada model with delayed feedback}
\label{sec:background model}

The bifurcation analysis of the Yamada model in \cite{Dubbeldam_99} and \cite{Krauskopf_book_chapter}, demonstrated, for $\kappa =0$, the existence of an equilibrium:
\begin{equation}
E_1: (G,Q,I) = (A,B,0).
\label{eq:non lasing equilibrium}
\end{equation}
Regardless of the other parameters values, this equilibrium has zero-intensity, and thus corresponds physically to the non-lasing or off solution.
In presence of feedback, two other equilibria $E_2$ and $E_3$ appear for a specific value of the feedback strength $\kappa$ through a saddle-node bifurcation \cite{Krauskopf_book_chapter}. Its locus is given by:
\begin{equation}
\kappa_S = \frac{-aA+a-B-1+2\sqrt{aAB}}{a-1}.
\label{eq:kappa_S}
\end{equation}
If $\kappa<\kappa_S$, then $E_1$ is the only equilibrium of system (\ref{eq:Yamada_feedback}). Conversely, if $\kappa>\kappa_S$, both $E_1$, $E_2$ and $E_3$ are solutions of the system. $E_2$ is a saddle, $E_3$  is a node, and they are given by:
\begin{equation}
E_2: (G,Q,I) = \left(\frac{A}{1+I_+},\frac{B}{1+aI_+},I_+\right),
\end{equation}
\begin{equation}
E_3: (G,Q,I) = \left(\frac{A}{1+I_-},\frac{B}{1+aI_-},I_-\right),
\end{equation}
where
\begin{equation}
I_\pm= \frac{-aA+B+a+1-a\kappa-\kappa}{2a(\kappa-1)} \pm \frac{\sqrt{(aA-B-a-1+a\kappa-\kappa)^2 - 4a(\kappa-1)(A-B-1+\kappa)}}{2a(\kappa-1)}.
\label{eq:Ipm}
\end{equation}
The equilibria $E_2$ and $E_3$ display constant, non-zero intensity, and thus correspond physically to continuous-wave emission. 
As demonstrated in  \cite{Krauskopf_book_chapter}, $E_2$ undergoes a transcritical bifurcation when the feedback strength $\kappa$ is further increased to the value $\kappa_T$:
\begin{equation}
\kappa_T = 1-A+B. 
\label{eq:kappa_T}
\end{equation}
At this point, $E_2$ exchanges stability with $E_1$. Meanwhile, the value of $I_+$ given by (\ref{eq:Ipm}) becomes negative: $E_2$ enters the half-plane with $I<0$, and this solution is thus mathematically stable, but not physically relevant anymore. More precisely, as long as $\kappa<\kappa_T$, the equilibrium $E_1$ is attracting and $E_2$ is of saddle-type, with positive intensity. Conversely, $E_1$ is of saddle type for $\kappa>\kappa_T$. This value $\kappa_T$, which directly depends on the value of the pump parameter $A$, thus corresponds to the laser's threshold.
Other bifurcations of equilibria, as well as bifurcations of periodic solutions, are found in system (\ref{eq:Yamada_feedback}). However, they cannot be calculated analytically, and one has to use dedicated continuation software for DDEs \cite{Engelborghs_02,Engelborghs_01,Sieber_biftool}.


\section{Bifurcation study in the ($\tau,\kappa$)-plane}
\label{sec:diag tau kappa}

We now perform a bifurcation study of equations (\ref{eq:Yamada_feedback}) in the $(\tau,\kappa)$-plane of feedback parameters, for two different values of the pump parameter $A$, where for $\kappa = 0$ the laser is excitable or has a unique non-lasing equilibrium, respectively. To investigate the influence of the optical feedback, we consider the fixed values $B = 5.8$, $a=1.8$ and $\gamma=0.04$, as in \cite{Krauskopf_book_chapter}.

\subsection{Case of the SLSA in the excitable regime}
\label{subsec:small tau}

\begin{figure}[ht!]
\begin{center}
\includegraphics[width=0.98\textwidth]{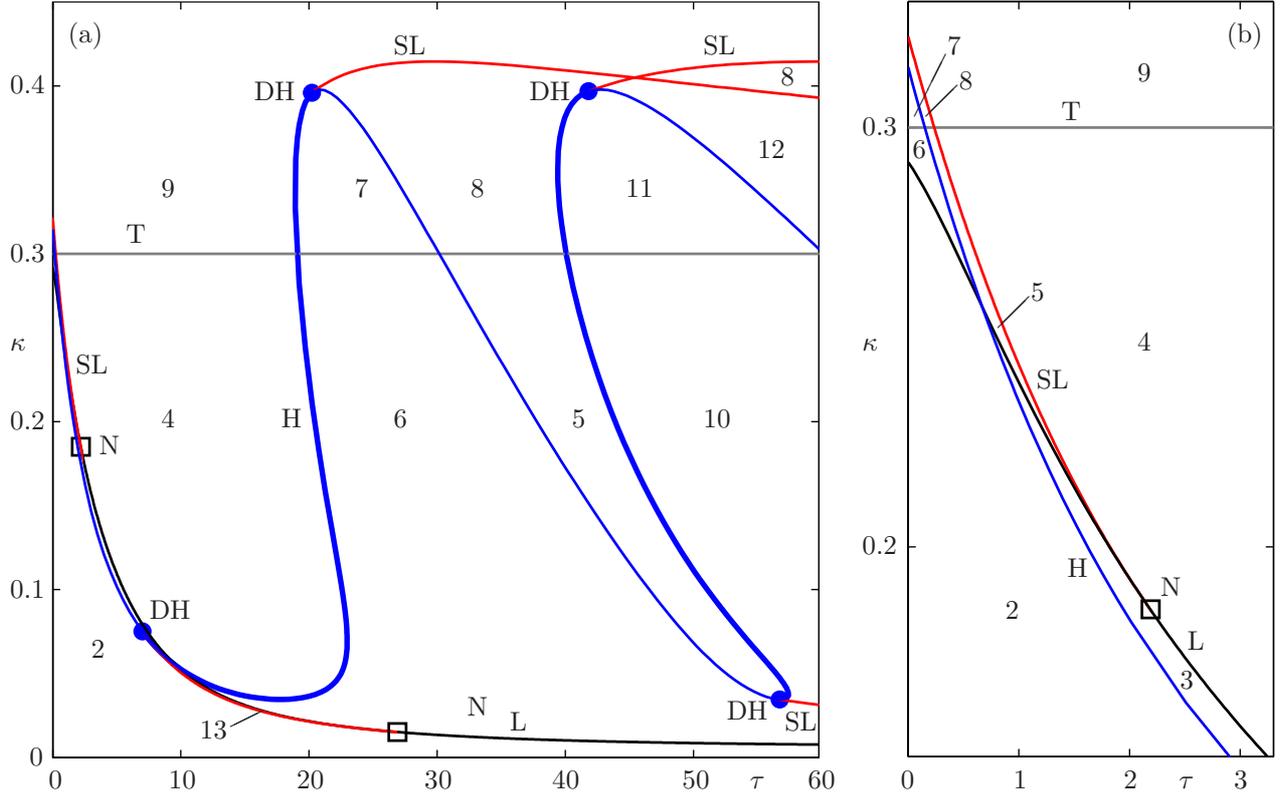}
\caption{Bifurcation diagram in the $(\tau,\kappa)$-plane for A = 6.5 (a). Panel (b) is an enlargement for small values of the delay $\tau$. Displayed are curves H of Hopf bifurcation (blue), T of transcritical bifurcation (grey), L of homoclinic loop (black) and SL of saddle-node bifurcations of periodic orbits (red). The labels DH and N correspond to points of degenerate Hopf and neutral saddle, respectively. Along bold parts of the curve H the Hopf bifurcation is supercritical. Numbered regions correspond to phase portraits in figure \ref{fig:phase portraits A6.5}.}
\label{fig:diag bif A6.5}
\end{center}
\end{figure}

\begin{figure}[h!]
\begin{center}
\includegraphics[width=0.9\textwidth]{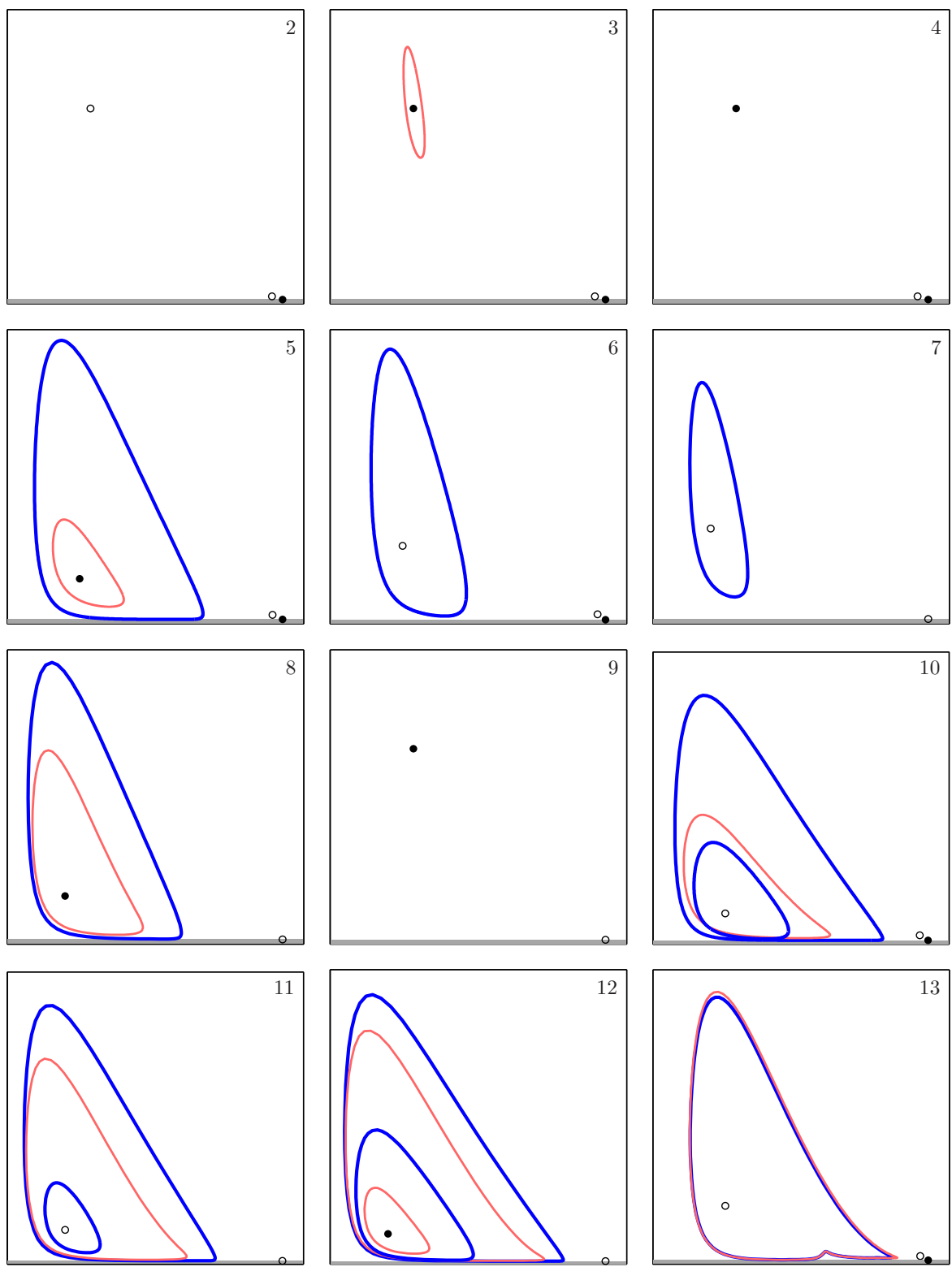}
\caption{Projection onto the $(G,I)$-plane of the phase portraits corresponding to regions of the bifurcation diagram numbered in figure \ref{fig:diag bif A6.5}(b), for $A=6.5$. Black dots and open dots are stable equilibria and saddle equilibria, respectively; blue curves are stable periodic orbits and red curves are saddle-type periodic orbits. Values of $\kappa$ and $\tau$ for each panels are given in appendix \ref{sec:appendix}.}
\label{fig:phase portraits A6.5}
\end{center}
\end{figure}

We first consider the case of a fixed pump parameter $A=6.5$. In this configuration, the laser without feedback is excitable \cite{Dubbeldam_99}, which corresponds to the phase portrait 2 in figure \ref{fig:phase portraits A6.5}. Figure \ref{fig:diag bif A6.5}(b) summarizes the main results of the bifurcation analysis performed in \cite{Krauskopf_book_chapter} for this configuration. The whole bifurcation diagram has been recomputed and includes new information, with extra curves representing the bifurcations of periodic solutions. 

Each curve of this two-parameters bifurcation diagram corresponds to the locus of a specific bifurcation, and the different curves thus divide the ($\tau,\kappa$)-plane into regions with qualitatively different dynamics. An overview of these different dynamics is provided in figure \ref{fig:phase portraits A6.5}, which represents two-dimensional projections, onto the $(G,I)$-plane, of the phase portraits associated with the different regions. Such a projection is relevant because, for small values of the delay, the dynamics effectively takes place in a two-dimensional attracting surface \cite{Krauskopf_book_chapter}. It is worth noting that  the numbering of the different regions and phase portraits is consistent with that in \cite{Dubbeldam_99} and \cite{Krauskopf_book_chapter}, the only phase portrait not reproduced here is that in region 1, where the equilibrium $E_1$ is a global attractor (which is not actually found in the bifurcation diagram in figure \ref{fig:diag bif A6.5}).

Overall, we find the following bifurcation curves and points of system (\ref{eq:Yamada_feedback}) (see, for example, \cite{Kuznetsov} for definitions and properties of these bifurcations):
\begin{itemize}
\item[T]: a curve of transcritical bifurcation, located at $\kappa_T$, where the equilibria $E_1$ and $E_2$ exchange stability. This corresponds, for example, to the transition between phase portraits 6 and 7 in figure \ref{fig:phase portraits A6.5}.
\item[S]: a curve of saddle-node bifurcation, located at $\kappa_S$, where the equilibria $E_2$ and $E_3$ bifurcate. This corresponds, for example, to the transition between phase portrait 1 where $E_1$ is a global attractor, and phase portrait 4.
\item[H]: curves of Hopf bifurcation, where the equilibrium $E_3$ changes stability and a periodic solution of small amplitude is created or disappears. The periodic solution is attracting when the bifurcation is supercritical (bold parts of curve H in figure \ref{fig:diag bif A6.5}(a)), and it is of saddle type when the Hopf bifurcation is subcritical. This corresponds, for example, to the transition between phase portraits 7 and 8 in figure \ref{fig:phase portraits A6.5}. 
\item[L]: a curve of homoclinic loop to saddle equilibrium $E_2$. Crossing this curve, a periodic orbit of very large period is either created or disappears. This is illustrated in figure \ref{fig:phase portraits A6.5} by the transition between phase portraits 3 and 5.
\item[SL]: curves of saddle-node bifurcations of periodic orbits, where two periodic orbits collide and disappear (or are created). An example is the transition between the phase portraits 8 and 12.
\item[TR]: curves of torus bifurcations of periodic orbits, where a periodic orbits has two complex conjugate Floquet multipliers crossing the unit circle, and an invariant torus bifurcates. This corresponds, for example, to the transition between phase portraits 14 and 15.
\item[DH]: degenerate Hopf bifurcation points, where the Hopf bifurcation changes criticality and a curve SL emerges. 
\item[N]: neutral saddle points along the homoclinic loop curve L, where the periodic orbit associated to the homoclinic loop changes stability, and a curve SL emerges. 
\item [FH]: Fold-Hopf bifurcation points, lying at the tangential intersection of the curve S and the curve H, where the system linearized around the equilibrium $E_3$ has a zero eigenvalue and two complex conjugate, purely imaginary eigenvalues.
\item[BT]: a Bogdanov-Takens point BT, lying along the curve S, where the system linearized around the equilibrium $E_3$ has a zero eigenvalue of multiplicity two. Generically, a curve H of Hopf bifurcation and a curve L of homoclinic loop emerge from this point. The periodic orbit arising from the Hopf bifurcation undergoes the homoclinic bifurcation.
\item [HH]: Hopf-Hopf bifurcation points, where two curves of Hopf bifurcations intersect, or a single Hopf bifurcation curve self-intersects. The system has two pairs of complex-conjugate, purely imaginary eigenvalues. Curves TR of torus bifurcation can emerge from these points.
\end{itemize}
Note that the saddle-node bifurcation of equilibria involving $E_2$ and $E_3$ does not occur in figure \ref{fig:diag bif A6.5}: for $A=6.5$ as considered in this section, the value $\kappa_S$ defined by equation (\ref{eq:kappa_S}) is negative, and $E_1$, $E_2$ and $E_3$ consequently all exist in the physically relevant half-plane with $\kappa>0$.

Figure \ref{fig:diag bif A6.5}(a) shows the precise locations of the curves of saddle-node bifurcations of limit cycles, which have now be continued numerically with DDE-Biftool \cite{Engelborghs_02,Sieber_biftool}. As theory predicts, they emerge from degenerate Hopf points DH and from codimension-two points N, where the Hopf bifurcations change criticality and the periodic orbit involved in the homoclinic loop changes criticality, respectively. 
Panel (b) represents an enlargement of the bifurcation diagram for low values of $\tau$. In this region, the curves H and L are found to cross each other twice, near $\tau =  0.67$  and $\tau = 9.8$. 
The criticality of L changes twice at codimension-two points N located at $\tau=2.2$ and $\tau = 26.9$. The curve SL emerging from the first point N approaches the limit $\kappa=0$, while the curve SL emerging from the second point N connects to the degenerated Hopf bifurcation point DH located near $\tau=7$. Between its two boundaries, this curve SL is just below the curve L  and stays very close to it. Together with curves H and L, the curve SL bounds a very small region 13 of the bifurcation diagram, whose phase portrait is represented in figure \ref{fig:phase portraits A6.5}. In this phase portrait, both the fact that the two periodic orbits are almost superimposed, and really close to the saddle equilibrium $E_2$, clearly highlights the proximity of both the saddle-node bifurcation of limit cycles and the homoclinic bifurcation. Note that phase portrait 13 highlights an additional small intensity peak along both periodic orbits, when the intensity is close to zero. This can be explained by the detuning between the delay $\tau$ and the period $T$: approaching the homoclinic bifurcation curve L, the period becomes much larger than the delay, and the second peak is thus due to a secondary excitation by the feedback. However, this feedback re-excitation occurs during the refractory period, when the gain has not entirely recovered: the next high-amplitude pulse cannot be triggered, and the returning pulse only results in a small increase and relaxation of the intensity.

Even for relatively small values of the delay, the bifurcation diagram \ref{fig:diag bif A6.5}(a) highlights new feedback-induced dynamics. This can be seen clearly when one considers the different dynamics encountered along a cross-section of the diagram, for a fixed value of the feedback strength $\kappa = 0.38$. For small values of the delay ($\tau<19.3$), in region 9, two equilibria exist and are physically relevant: $E_3$ is attracting, and $E_1$ is unstable.
When the delay is increased, the Hopf bifurcation curve H is crossed to enter region 7: the bifurcating equilibrium $E_3$ becomes unstable and a small-amplitude periodic solution is created; see figure \ref{fig:phase portraits A6.5} for the phase portrait. Increasing the delay further, the curve H is crossed a second time and $E_3$ undergoes a second (subcritical) Hopf bifurcation: entering region 8, a new periodic orbit of saddle type is created, while $E_3$ becomes stable again. The same process repeats when regions 11 and 12 are entered: the Hopf bifurcation curve is crossed two additional times, leading each time to the birth of a new periodic orbit, which is alternately stable and of saddle type. Different stable periodic solutions with different amplitudes thus coexist in regions 10, 11 and 12. Physically, this means that the laser can display different qualitative and quantitative behaviors for the same parameter values: depending only on the initial conditions, the output is either a light beam with constant intensity, or pulse-like self-oscillations with different amplitudes and periods.

\subsubsection{Extending the range of the delay}
\label{subsub:A6.5 large delays}

\begin{figure}[h!]
\begin{center}
\includegraphics[width=\textwidth]{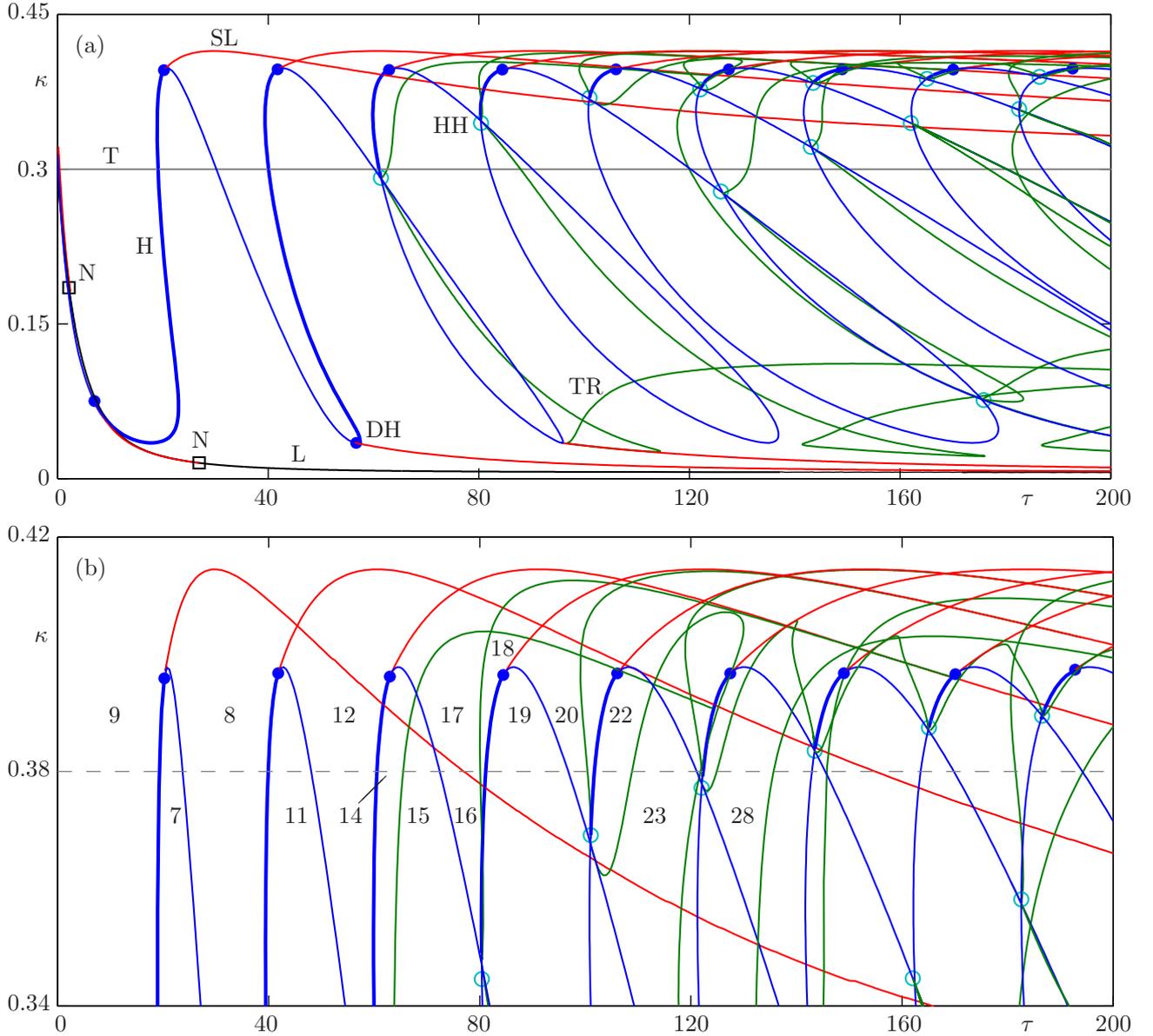}
\caption{Extended bifurcation diagram in the $(\tau,\kappa)$-plane, for $A=6.5$ and $\tau \in [0, 200]$. Displayed are curves H of Hopf bifurcation (blue), T of transcritical bifurcation (grey), L of homoclinic loop (black), SL of saddle-node bifurcations of periodic orbits (red) and TR of torus bifurcations (green). The labels DH, N and HH correspond to points of degenerate Hopf bifurcations, neutral saddle and Hopf-Hopf bifurcations, respectively. Bold parts of the curve H indicate that the Hopf bifurcation is supercritical. Panel (b) is an enlargement near $\kappa = 0.38$, where some of the regions crossed while increasing the delay for $\kappa = 0.38$ are numbered, the selected phase portraits are displayed in figures \ref{fig:phase portraits A6.5}, \ref{fig:phase portraits A6.5 bis} and \ref{fig:stable QP sol A6.5}.}
\label{fig:diag bif A6.5 large}
\end{center}
\end{figure}

\begin{figure}[h!]
\begin{center}
\includegraphics[width=\textwidth]{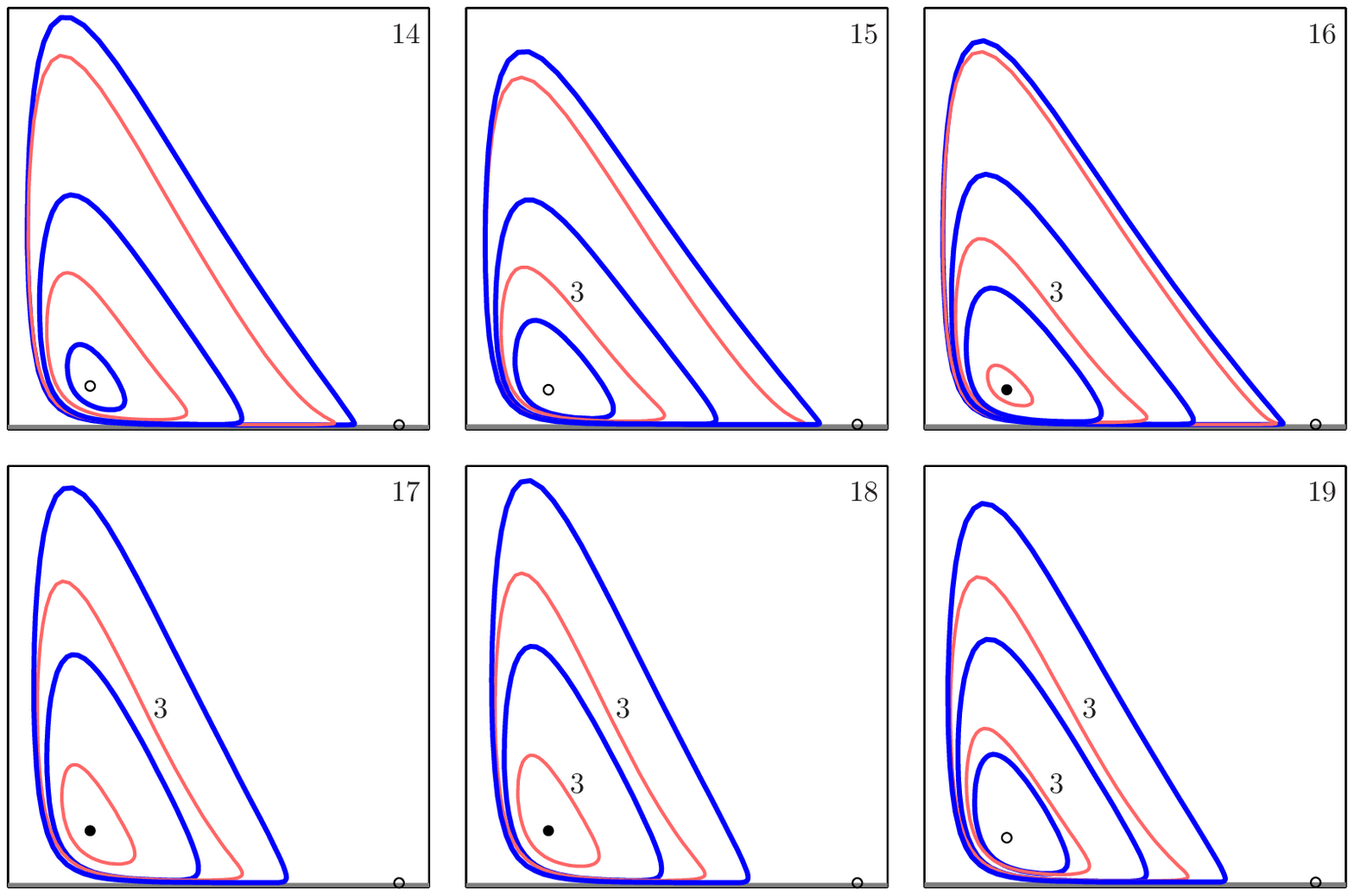}
\caption{Projection onto the (G,I)-plane of the phase portraits in regions 14 -- 19 of the bifurcation diagram in figure \ref{fig:diag bif A6.5 large}. Black dots and open dots are stable equilibria and saddle equilibria, respectively. Blue and red curves are stable and saddle-type periodic orbits, respectively. The numbers near saddle-type periodic orbits refer to the number of Floquet multipliers outside the unit circle, when this number is larger than one. Parameters values are A=6.5 and $(\kappa,\tau)=(0.38,63)$ in panel 14; $(\kappa,\tau)=(0.38,68)$ in panel 15; $(\kappa,\tau)=(0.38,75)$ in panel 16; $(\kappa,\tau)=(0.38,79)$ in panel 17; $(\kappa,\tau)=(0.38,80.5)$ in panel 18; and $(\kappa,\tau)=(0.38,90.5)$ in panel 19.}
\label{fig:phase portraits A6.5 bis}
\end{center}
\end{figure}

Although the values of $\tau$ considered in figure \ref{fig:diag bif A6.5}(a) can seem large compared to the internal time scale of the SLSA, these values remain small from an experimental point of view. Indeed, optical paths of up to several meters are commonly used for the feedback loop in the experiment \cite{Felix_master_thesis}. 
Figure \ref{fig:diag bif A6.5 large}(a) presents an extended bifurcation analysis for larger values of $\tau$.
It is a well-known property of delay systems, that periodic solutions reappear infinitely many times for different values of the delay \cite{Yanchuk}. The consequence here is that the curve H starts to self-intersect for sufficiently large values of the delay. Each of these intersection points is a Hopf-Hopf bifurcation point HH, where the system has two pairs of complex-conjugate, purely imaginary eigenvalues. 
From each of these codimension-two points, two curves TR of torus bifurcation of periodic orbits can emerge. When a periodic orbit undergoes a torus bifurcation, two complex-conjugate Floquet multipliers cross the unit circle, and a (stable or saddle-type) invariant torus bifurcates from the periodic solution \cite{Roose2007continuation}.
These additional bifurcation curves make the bifurcation diagram in figure \ref{fig:diag bif A6.5 large}(a) increasingly complicated as the delay is increased. In practice, it becomes impossible to map out the dynamics in each of the many different regions of the ($\tau,\kappa$)-plane. 
Considering, as before, a cross-section of the bifurcation diagram for a fixed value of $\kappa$ highlights new feedback-induced dynamics. Figure \ref{fig:diag bif A6.5 large}(b) represents an enlargement of the bifurcation diagram in panel (a) for $\kappa = 0.38$, and figure \ref{fig:phase portraits A6.5 bis} presents some of the additional phase portraits encountered along this section while increasing $\tau$.

Starting from region 12 where we stopped in section \ref{subsec:small tau}, increasing $\tau$ leads to the crossing of the Hopf bifurcation curve H to enter region 14, where we find the coexistence of three stable periodic solutions. When the torus bifurcation curve TR, which separates regions 14 and 15, is crossed one of the unstable periodic orbits existing in region 14 undergoes a torus bifurcation: in region 15 it thus has two additional Floquet multipliers outside the unit circle. Numbers near unstable periodic orbits in figure \ref{fig:phase portraits A6.5 bis} indicate the number of Floquet multipliers outside the unit circle, when this number is larger than one. It is not possible with DDE-Biftool to determine the criticality of the torus bifurcation; this thus has to be investigated through time-domain simulations, it will be discussed in section \ref{subsub:torus}.
The transition from region 15 to 16 is of the same nature as the transition between regions 11 and 12 previously described: a new unstable periodic orbit is thus observed in region 16. Up to this point, increasing the delay for the fixed value $\kappa = 0.38$ has led to coexistence of more and more stable periodic orbits. However, when the curve SL of saddle-node bifurcation of periodic orbits is crossed to enter region 17, the stable and the unstable periodic orbits with the largest amplitudes collide and disappear. In region 17, the system thus goes back to multistability between only two periodic orbits and equilibrium $E_3$. Transition between regions 17 and 18, through the crossing of a torus bifurcation curve, is very similar to the transition from region 14 to region 15. The curve H is crossed to enter region 19, and the phase portrait thus again shows the coexistence of three different stable periodic orbits.

\subsubsection{Multistability: features of periodic orbits}
\label{subsub:multistab}

\begin{figure}[h!]
\begin{center}
\includegraphics[width=\textwidth]{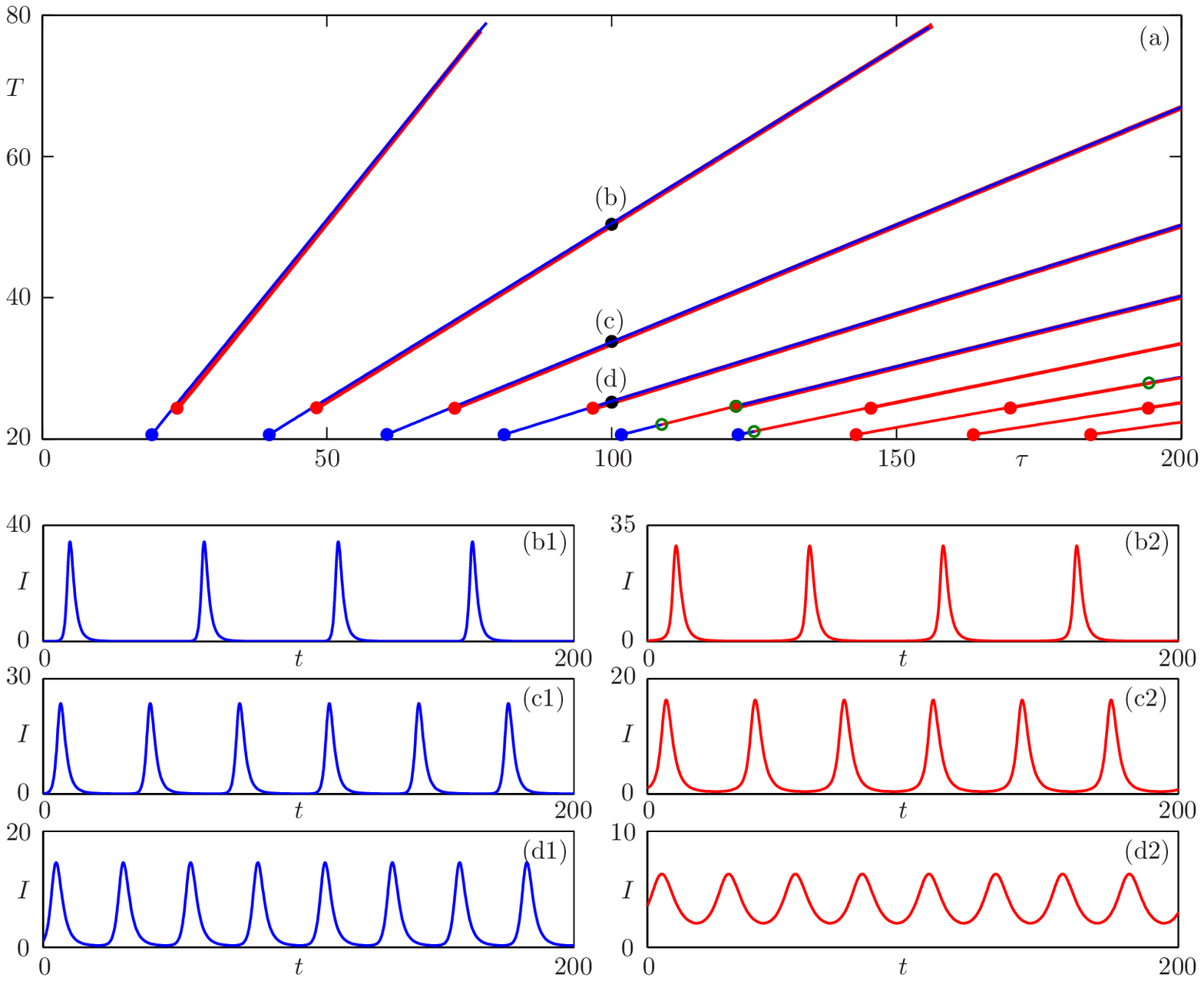}
\caption{Families of periodic orbits of (\ref{eq:Yamada_feedback}) for $A=6.5$ and $\kappa= 0.38$: evolution in $\tau$ of the period $T$ along the different periodic solution branches (a). Stable and unstable parts of the periodic solutions branches are represented in blue and red, respectively; blue and red dots are supercritical and subcritical Hopf bifurcations, respectively, and open green dots are torus bifurcations. For the six periodic orbits coexisting at $\tau = 100$, time series of the intensity $I$ are represented in panels (b1), (c1), (d1) for the stable solutions and in panels (b2), (c2) and (d2) for the unstable solutions.}
\label{fig:time series multistab A6.5 tau 100 kappa 0.38}
\end{center}
\end{figure}

Figure \ref{fig:time series multistab A6.5 tau 100 kappa 0.38} represents, for fixed $\kappa=0.38$, the evolution of the period $T$ of periodic orbits with respect to $\tau$ in panel (a) and the time series of periodic orbits at $\tau = 100$ in panels (b), (c) and (d). For the parameter set considered here and for all the branches of periodic solutions, $T$ is never smaller than $20.6$ at the Hopf bifurcation point, and is always $T\approx78.7$ at the saddle-node bifurcation constituting the other end of the branch. 
Physically, these values are related to the recovery and saturation times of the gain and absorber media; that is to say, it is determined by the value of  $\gamma$ in equations (\ref{eq:Yamada_feedback}).

In figure \ref{fig:time series multistab A6.5 tau 100 kappa 0.38}(a), the period evolves practically linearly with the delay $\tau$, with different slopes depending on the considered periodic solution branch.
Along the first pair of periodic orbit branches -- one attracting which is born at $\tau=19.3$, and one of saddle type which is born at $\tau=23.7$ -- the period stays really close to the value of the delay. In other words, the period $T$ evolves with a slope of 1 with respect to the delay $\tau$ along these branches. 
The period of the two following periodic orbits -- one attracting and one of saddle-type, which are born at $\tau= 39.9$ and at $\tau=48.1$, respectively -- stays close to half the delay along their branches.  
In the same way, the next pair of periodic orbits have, along their corresponding branches, a period close to a third of the delay, and so on. This corresponds to one, two or three pulses per roundtrip of the external feedback loop.

For a fixed value of the delay of $\tau=100$, we now compare the time series corresponding to the different periodic orbits. For the considered value of $\tau$, figure \ref{fig:time series multistab A6.5 tau 100 kappa 0.38}(a) shows the coexistence of three pairs of periodic orbits (b), (c) and (d), each pair consisting of an attracting and a saddle-type periodic solution.
Time series of the intensity $I$  are shown in panels (b1), (c1) and (d1) for the stable periodic orbits, and in panels (b2), (c2) and (d2) for the saddle periodic orbits. 
Beyond the differences in terms of the repetition rate of the pulses already discussed above, this highlights significant differences in the temporal shape of the laser output. Panels (b1) and (b2) show high-amplitude, narrow pulses, with low repetition rates of about $\tau$. Compared to point (b), point (c) in figure \ref{fig:time series multistab A6.5 tau 100 kappa 0.38}(a) is closer to the respective Hopf point where the periodic solution branches are born. Compared to times series (b1) and (b2), the pulse-like oscillations in time series (c1) and (c2) shows wider pulses, with smaller amplitude and higher repetition rate of about $\tau/2$. In the same way, at point (d) we are closer to the Hopf bifurcation points where the considered periodic orbits are born than for point (c). In particular, the periodic orbit of saddle-type is born at the Hopf point at $\tau =96.7$. The corresponding time series for $\tau = 100$ in panel (d2) thus shows, as expected, small-amplitude and almost sinusoidal oscillations. The stable periodic orbit is born at $\tau =81.1$; at point (d), we are thus further away from the Hopf bifurcation than in the case of the saddle-type periodic orbit. The corresponding time series in panel (d1) is therefore more pulse-shaped, but the pulses are wider, and their amplitude smaller than for times series (b1) and (c1).


\begin{figure}[h!]
\begin{center}
\includegraphics[width=0.95\textwidth]{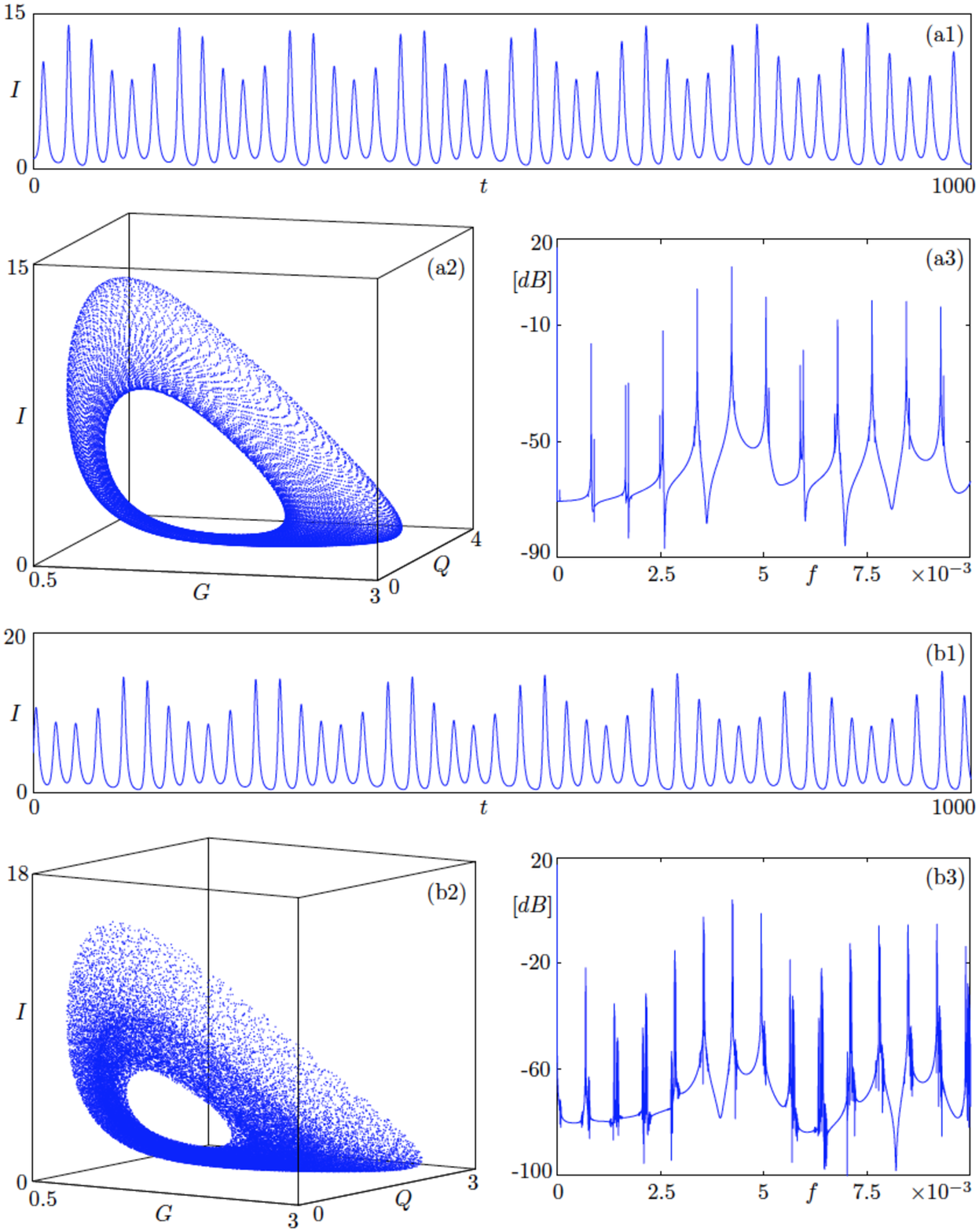}
\caption{Stable dynamics on a torus in regions 23 and 28 of figure \ref{fig:diag bif A6.5 large}, for $A=6.5$ and  $\kappa=0.38$, with $\tau = 117$ (a) and with $\tau = 140$ (b). Shown are the intensity time series in panels (a1) and (b1), the projections onto the $(G,Q,I)$-space in panels (a2) and (b2), and the spectra (in dB) in panels (a3) and (b3).}
\label{fig:stable QP sol A6.5}
\end{center}
\end{figure}

\subsubsection{Bifurcating tori}
\label{subsub:torus}

So far, we only considered multistability between periodic orbits and equilibria. We now investigate transitions through torus bifurcations curves TR, which appear for sufficiently large $\tau$, for example, between regions 22 and 23 and at the transition into region 28. In both cases, the periodic orbit undergoing the torus bifurcation is stable before the bifurcation (\textit{i.e.} for smaller values of the delay, along the cross section defined by $\kappa=0.38$), and thus becomes unstable in the torus bifurcation. Because invariant tori cannot be continued for delay systems, we run time-domain simulations with dde23 \cite{dde23}, starting from the bifurcating periodic orbit as initial condition. 

 Figure \ref{fig:stable QP sol A6.5} represents, in panels (a) and (b), the stable tori in regions 23 and 28, respectively.
In both cases, the time series in panels (a1) and (b1) suggest either quasiperiodic oscillations or high-period locked periodic solutions on a torus.
Projections of these solutions onto the physical $(G,Q,I)$-space in panels (a2) and (b2) highlight the two associated invariant tori. The spectra of both time series, in panels  (a3) and (b3), are clearly non harmonic, and both of them are composed of discrete peaks at frequencies equal to linear combinations of two base frequencies. Altogether, figure \ref{fig:stable QP sol A6.5} confirms the quasiperiodic feature of the oscillations as taking place on an invariant torus. 

Both in regions 23 and 28, the system thus displays multistability between a stable invariant torus and different periodic orbits (three in region 23, and four in region 28). The stable invariant torus in region 23 disappears when the same torus bifurcation curve TR is crossed again, which occurs at $\tau \approx  121.8$ for $\kappa =0.38$; see figure \ref{fig:diag bif A6.5 large}(b). At this point, the bifurcating periodic orbit regains stability. For the considered values of $\tau$, we do not observe a coexistence of the two stable invariant tori. However, the increasing complexity of the bifurcation diagram for large values of $\tau$ suggests that the coexistence of several stable invariant tori and stable periodic orbits is to be expected. 
The dynamics of the Yamada model without delay has been shown to take place in a globally attracting two-dimensional surface \cite{Dubbeldam_99}. The existence of stable tori clearly shows that the dynamics of the SLSA with feedback is no longer two-dimensional for sufficiently large $\tau$.



\subsection{Case of the SLSA in a non-excitable off state}
\begin{figure}[h!]
\begin{center}
\includegraphics[width=\textwidth]{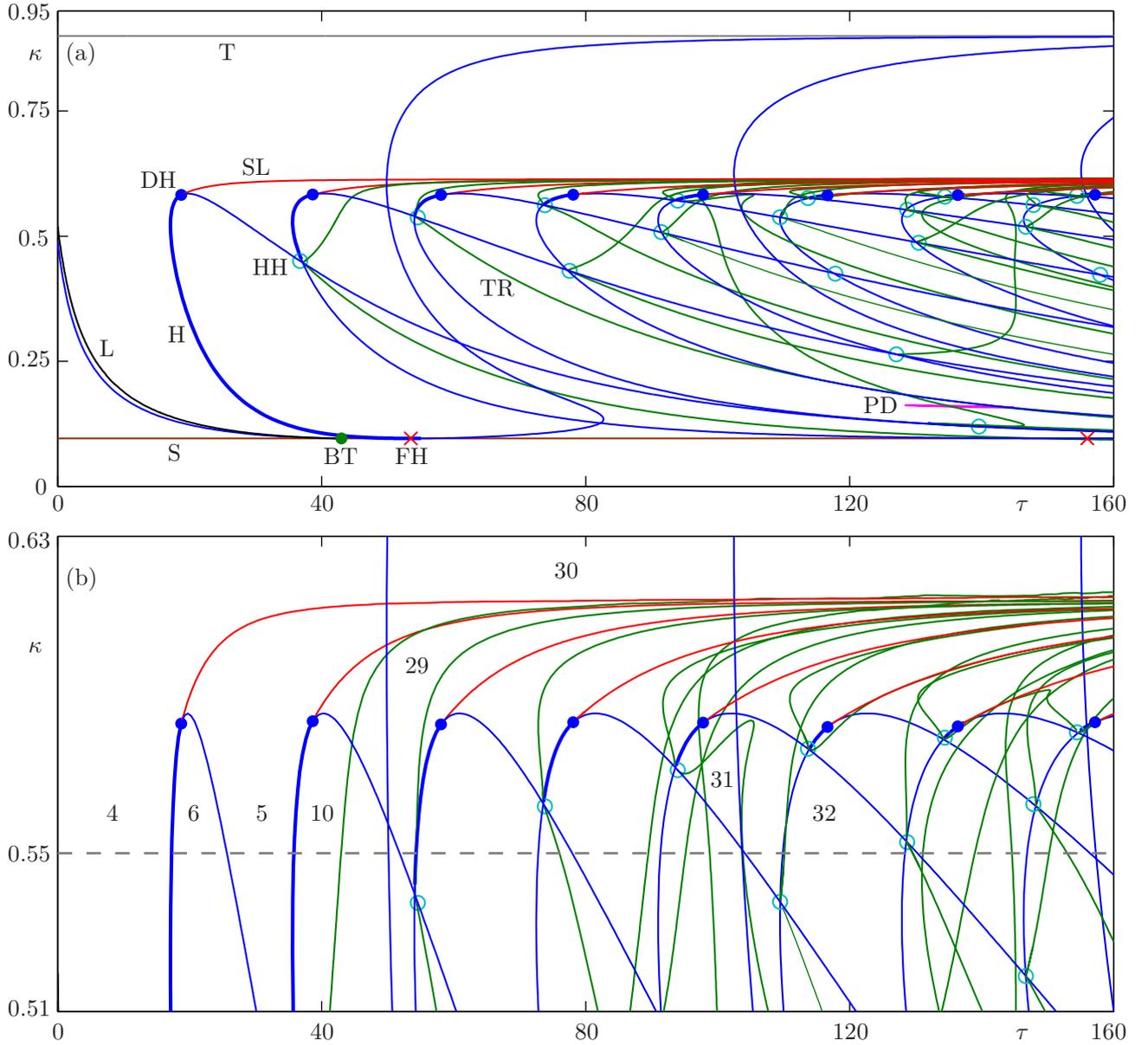}
\caption{Bifurcation diagram in $(\tau,\kappa)$-plane, for A = 5.9 (a). Panel (b) is an enlargement near $\kappa = 0.57$. Displayed are curves H of Hopf bifurcations (blue), T of transcritical bifurcation (grey), S of saddle-node bifurcation (brown), L of homoclinic loop (black), TR of torus bifurcations (green), SL of saddle-node bifurcations of periodic orbits (red) and PD of period-doubling bifurcations (pink). The labels HH, DH, BT and FH correspond to points of Hopf-Hopf, degenerate Hopf, Bogdanov-Takens and Fold-Hopf bifurcations, respectively. Along bold parts of the curve H the Hopf bifurcation is supercritical. Some regions are numbered, and corresponding phase portraits are displayed in figures \ref{fig:phase portraits A6.5} and \ref{fig:PP A5.9}.}
\label{fig:diag bif A5.9}
\end{center}
\end{figure}

From a practical point of view, the influence of the pump parameter $A$ is particularly interesting, because it can easily be changed experimentally.
We now consider the influence of the delayed optical feedback on the dynamics of the Yamada model for a fixed value of $A=5.9$. In this case, the SLSA without feedback is in a non-excitable regime: the non-lasing equilibrium $E_1$ is the unique steady-state \cite{Dubbeldam_99}.
The influence of the delayed optical feedback on this phase portrait has been investigated in \cite{Krauskopf_book_chapter}, but for small values of $\tau$ only. 
For the same configuration, figure \ref{fig:diag bif A5.9} (a) represents an extended bifurcation diagram in the $(\tau,\kappa)$-plane.
The curve S, where equilibria $E_2$ and $E_3$ bifurcate, and the curve T, where $E_1$ and $E_2$ exchange stability, are located at $\kappa_S \approx 0.0958$ and $\kappa_T = 0.9$, respectively; see section \ref{sec:background model}. Contrary to the excitable case in section \ref{sec:diag tau kappa}, multiple disjoint curves H of Hopf bifurcation are found, which either involve $E_2$ or $E_3$. Moreover, codimension-two bifurcation points occur along the curve S:  two Fold-Hopf points FH, and a Bogdanov-Takens point BT; see section \ref{sec:diag tau kappa} for their description. A curve H and a curve L of homoclinic loop (involving the periodic orbit that appears at the Hopf bifurcation) emerge from BT, and they both end at the left limit $\tau=0$ of the bifurcation diagram. As before, curves TR emerge from Hopf-Hopf points HH where curves H intersect, and SL curves emerge from degenerate Hopf bifurcation points DH where a Hopf bifurcation change criticality. Note that now, the different curves SL tend to an asymptotical value of $\kappa$ when $\tau$ is increased: continuing one of the branch SL up to $\tau = 1000$ clearly shows that it remains located between $\kappa= 0.61$ and $\kappa = 0.62$.
These qualitative differences with the case of the SLSA in the excitable regime in figure \ref{fig:diag bif A6.5 large} has multiple consequences for the system's dynamics. Since curve S in figure \ref{fig:diag bif A5.9} is in the physically-relevant half-space $\kappa>0$, the equilibrium $E_1$ is the only stable solutions for $\kappa<\kappa_S$. Moreover, $E_1$ remains stable for $\kappa<\kappa_T$, that is to say, in most regions of the bifurcation diagram. Here, multistability thus involves not only pulsing regimes and the continuous wave solution $E_3$, but also the off-state $E_1$. 


\begin{figure}[h!]
\begin{center}
\includegraphics{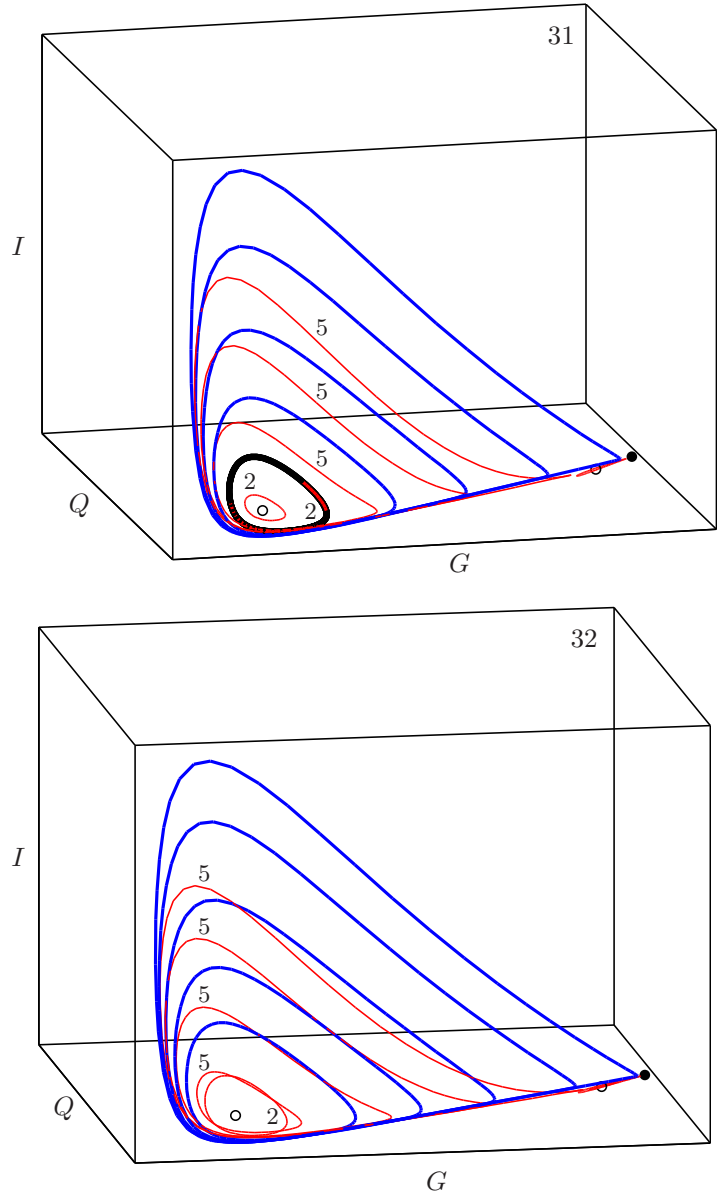}
\caption{Projection onto the (G,Q,I)-space of the phase portraits in regions 31 (top), and 32 (bottom) of figure \ref{fig:diag bif A5.9}. Black and open dots are stable and saddle equilibria, respectively. Blue and red curves are stable and saddle-type periodic orbits, respectively. The number of Floquet multipliers outside the unit circle is indicated near unstable periodic orbits when this number is larger than one.
In phase portrait 31, the black curve is a stable torus, which surrounds an unstable periodic orbit.}
\label{fig:PP A5.9}
\end{center}
\end{figure}

As before, the complexity of the bifurcation diagram in figure \ref{fig:diag bif A5.9}(a) makes it impossible in practice to map out the dynamics in all the different regions. However, the enlargement of the bifurcation diagram for the cross-section near $\kappa=0.57$, in panel (b), highlights some of the new dynamics.
Along this cross-section, no new dynamics are found for $\tau<40$, and the different regions correspond to phase portraits already observed for the case of the SLSA in the excitable regime; see figures \ref{fig:phase portraits A6.5} and \ref{fig:phase portraits A6.5 bis}. For the selected regions where new dynamics are found for $\tau>40$, figure \ref{fig:PP A5.9} represents the additional phase portraits. Namely, a curve H of Hopf bifurcation involving the equilibrium $E_2$ is crossed to enter region 29, and a small-amplitude periodic orbit is created around $E_2$, which can be seen in the phase portraits of regions 31 and 32.
For the range of parameters considered in figure \ref{fig:diag bif A5.9}(a), this Hopf bifurcation is subcritical, and the corresponding periodic orbit stays unstable when varying $\kappa$ and $\tau$. Additionally, no curve SL is crossed along the cross-section $\kappa=0.57$, as they all tend to an asymptotical value close to $\kappa = 0.62$. On the other hand, more and more curves H are crossed when increasing $\tau$, and more and more periodic orbits are thus created. As a consequence, the phase portraits in regions 31 and 32, in figure \ref{fig:PP A5.9}, show the coexistence of the stable non-lasing equilibrium $E_1$ with four and five stable periodic orbits, as well as five and six unstable periodic orbits, respectively. In contrast, no more than four co-existing stable periodic orbits were found, for the same range of $\tau$, for the case of the SLSA in the excitable regime. As discussed in section \ref{subsub:multistab}, the period $T$ of the different periodic orbits increases linearly with $\tau$. Here, this means that more and more stable periodic solutions are observed when $\tau$ is increased, which correspond to narrow, high-amplitude light pulses.
Moreover, a stable torus is born while entering region 31, which is computed by time-domain simulation and represented in black in figure \ref{fig:PP A5.9}(a). This torus being very thin, the corresponding pulsing solution only shows a very weak amplitude modulation. From a practical point of view, it can thus be easily mixed up with a periodic orbit. 

Although the introduction of a delayed optical feedback leads to qualitatively different behaviours for the SLSA in an excitable and a non excitable regime, several common features can be highlighted. In both cases, the bifurcation diagrams quickly becomes intricate when the feedback delay in increased beyond value of $\tau \leq 60$ studied in \cite{Krauskopf_book_chapter}, with many bifurcations occurring in small parameters regions. While the case of the SLSA in the non excitable regime corresponds to very simple dynamics, the introduction of delayed feedback leads to a complicated bifurcation diagram, even for relatively small delay times compared to the experimental configuration \cite{Felix_master_thesis}. In both cases considered so far, multistability considerably increases with the value of $\tau$, so that coexistence of several pulsing solutions is always observed when the delay is large enough. Finally, the systematic presence of torus bifurcations demonstrates that, in contrast to the system without delay  \cite{Dubbeldam_99}, the dynamics is not two-dimensional anymore from intermediate values of $\tau$. This also suggests that coexistence of several stable tori is to be expected for large values of the delay.


\section{Bifurcation study in the $(A,\kappa)$-plane}
\label{sec:bif study A kappa}

To study the influence of the feedback, we considered so far bifurcation diagrams in the $(\tau,\kappa)$-plane. Firstly to understand the influence of both feedback parameters, and secondly because the DDE for $\kappa>0$ is a regular perturbation of the ODE for $\kappa=0$, whose dynamics has been thoroughly studied in \cite{Dubbeldam_99}. 
The comparison of the bifurcation diagrams in the $(\tau,\kappa)$-plane for two values of the pump parameter A (in figures \ref{fig:diag bif A6.5 large} and \ref{fig:diag bif A5.9}) has shown that the bifurcation scenario strongly depends on $A$, not only quantitatively but also qualitatively.
We focus in this section on bifurcation diagrams of system (\ref{eq:Yamada_feedback}), in the plane of feedback strength $\kappa$ and pump parameter $A$. These are particularly relevant from a practical point of view: the pump A is the main control parameter of the laser, and $\kappa$ can easily be tuned with an attenuator. The delay time $\tau$, on the other hand, cannot be changed easily in an experiment.

\begin{figure}[h!]
\begin{center}
\includegraphics[scale=0.86]{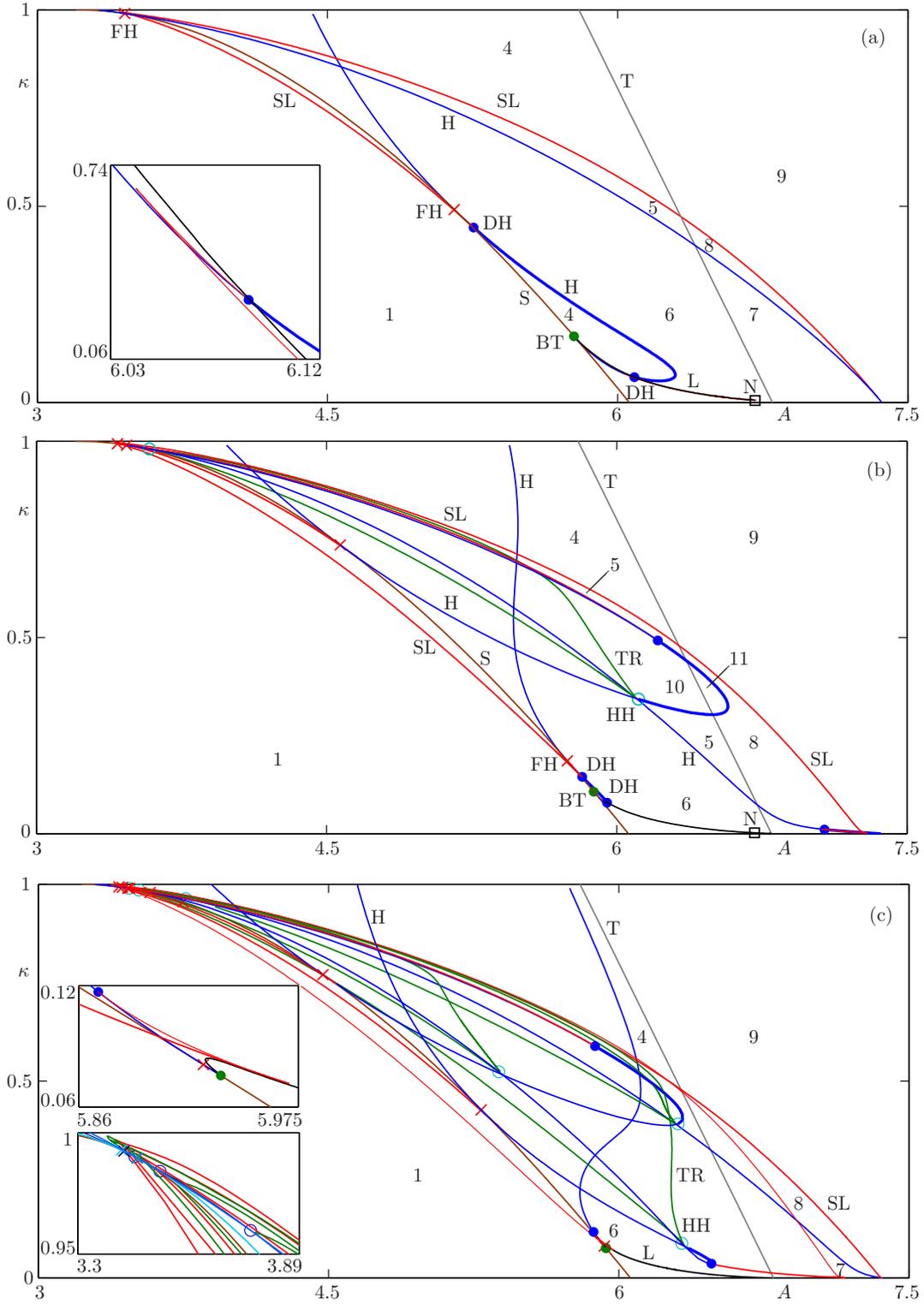}
\caption{Bifurcation diagrams in the $(A,\kappa)$-plane for $\tau = 25$ (a), $\tau = 40$ (b) and $\tau=58$ (c). Displayed are curves H of Hopf bifurcations (blue), T of transcritical bifurcation (grey), S of saddle-node bifurcation (brown), L of homoclinic loop (black), SL of saddle-node bifurcations of periodic orbits and TR of torus bifurcations.The labels HH, DH, BT, FH and N correspond to points of Hopf-Hopf bifurcations, degenerate Hopf bifurcations, Bogdanov-Takens bifurcation, Fold-Hopf bifurcations, and neutral saddle, respectively. Insets in panels (a) and (c) are enlargements near points of interest.}
\label{fig:bif diag A kappa tau 25 40}
\end{center}
\end{figure}

Figure \ref{fig:bif diag A kappa tau 25 40} shows the three bifurcation diagrams in the $(A,\kappa)$-plane for $\tau = 25$ in panel (a), $\tau = 40$ in panel (b) and $\tau = 58$ in panel (c). For small value of $\tau$, in panel (a), the bifurcation scenario remains relatively simple, and the dynamics in all the different regions can easily be mapped. For an intermediate value of $\tau = 40$, in panel (b), both the bifurcation diagram, and the sequence of bifurcations encountered when increasing A for a fixed value of $\kappa$, become more intricate. Nevertheless, it remains possible to identify the dynamics in the different regions, especially for reasonably low values of $\kappa$, considered in practice. On the other hand, when the delay is increased further, in panel (c), the bifurcation diagram has considerable complexity: many different bifurcation curves and regions with different dynamics are found in small parameter ranges, so that it is hard to see the details of the bifurcation scenario. 
Figure \ref{fig:bif diag A kappa tau 25 40} provides an additional and complementary representation of the increasing complexity of the dynamics when $\tau$ becomes larger, by illustrating how cross-sections of the bifurcation diagrams in figures \ref{fig:diag bif A6.5 large} and \ref{fig:diag bif A5.9} for fixed $\tau$, evolve with the pump parameter A. 

More precisely, for $\tau=25$ in figure \ref{fig:bif diag A kappa tau 25 40} (a), most of the phase portraits have already been observed for the system without feedback \cite{Dubbeldam_99}. A new feature, however, is the occurence of two cusp bifurcations of saddle-node bifurcations curves SL: one cusp is found close to $\kappa=1$, and the second one is highlighted in the inset of panel (a). Here, bistability occurs between equilibria and a single periodic orbit; however $\tau$ is not sufficiently large to observe multistability between several periodic orbits. We conclude that, although the sequence of bifurcations encountered when increasing $A$ depends on $\kappa$, some constant features are observed. Namely, the SLSA is necessarily in the off-state for small $A$ (region 1), and on the continuous-wave solution $E_3$ for large $A$ (region 9). In between these two regions, self-pulsations can occur in regions 5, 6, 7 and 8. However, all of them except for 7 show bistability between a pulsing solution and at least one equilibrium.  
From a practical point of view, this means that region 7 is the only one where self-pulsations are necessarily observed.

For the larger value of $\tau = 40$ in panel (b), the bifurcation diagram becomes more complex: the accumulation of both cusp points and curves H, SL, and TR around $(A,\kappa) = (3.5,1)$ makes it impossible in practice to distinguish between all the different regions. For this value of $\tau$, new dynamics is found compared to the case without feedback, including multistability between two different periodic orbits in regions 10 and 11.
As a consequence, for large values $\kappa>0.3$, the bifurcation scenario observed when increasing or decreasing A becomes more complicated: it includes jumps between different periodic solutions, hysteresis phenomena, and many different dynamics in a tiny parameter range. On the other hand, for small values $\kappa<0.3$, which are most often considered in experiments, no new dynamics is observed compared to the case without feedback. Moreover, the bifurcation sequence observed when increasing or decreasing A for fixed values of $\kappa$ is similar as that for $\tau=25$.

For the delay of $\tau = 58$ in panel (c), the bifurcation diagram clearly illustrates the increasing complexity of the system dynamics. Namely, an additional curve H is found, along which both points DH and HH occur. The emergence, from these points, of additional curves SL and TR, respectively, leads to a particularly intricate diagram. In particular, the bottom inset shows an enlargement near the point $(A,\kappa)=(3.5,1)$, where more and more bifurcation curves and codimension-two bifurcation points accumulate when $\tau$ is increased. In the same way, the top inset is an enlargement near the Bogdanov-Takens point BT: it clearly shows that, for this value of $\tau$, the bifurcation diagram becomes intricate even for small values of $\kappa$, with many bifurcation curves in a small parameters range. Some of the numerous new regions found in figure \ref{fig:bif diag A kappa tau 25 40} (c) show multistability between different pulsing solutions. It is possible to observe up to three stable periodic solutions, corresponding to light pulses of different amplitudes and repetition rates. On the other hand, in spite of the existence of many curves TR, no stable torus is found for this value of $\tau$.

All together, the bifurcation diagrams in the $(\tau, \kappa)$-plane and in the $(A, \kappa)$-plane show that the SLSA's dynamics becomes more and more complex as the delay $\tau$ is increased. Even for relatively small values of $\tau$, the feedback triggers complex bifurcation scenario, with a wealth of new interesting dynamics, including multistability of pulse-like periodic solutions with different amplitude and repetition rate. It is of practical importance that these new dynamics, which were first highlighted in the $(\tau, \kappa)$-plane, are also accessible in the $(A,\kappa)$-plane for small values of $\kappa$. Indeed, in an experiment, only values of $\kappa<0.5$ can usually be considered \cite{Felix_master_thesis}. This means that the new regimes described above can be observed in an experiment when $A$ is changed.


\section{Consequences of multistability}
\label{sec:multistability}

\begin{figure}[t!]
\includegraphics[width=\textwidth]{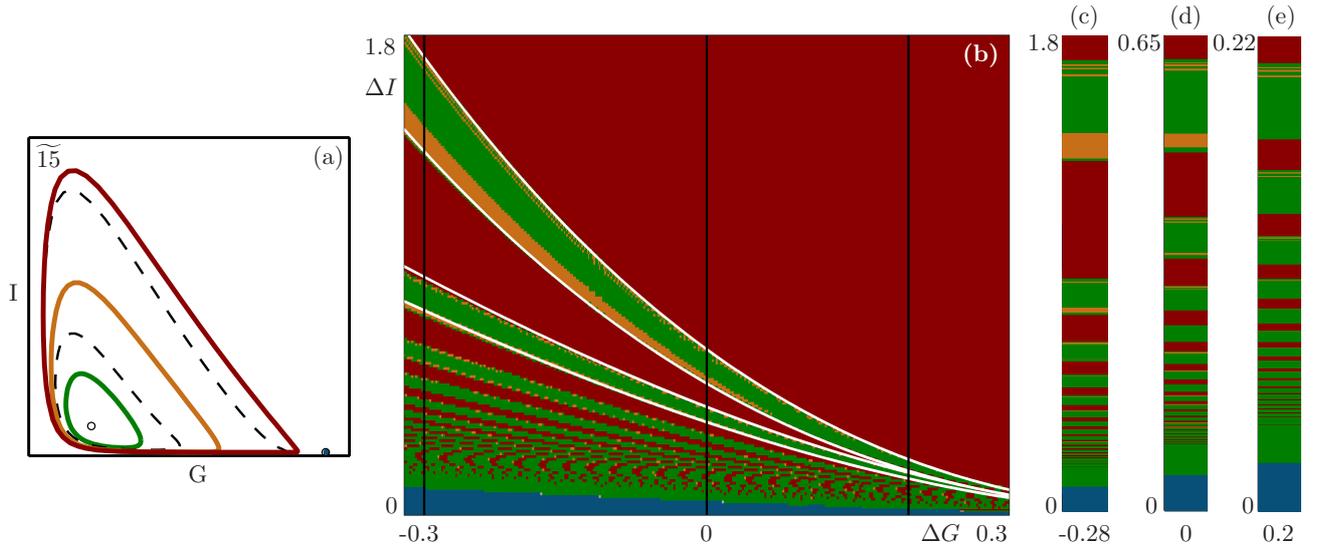}
\caption{Projection onto the (G,I)-plane of the phase portrait of system (\ref{eq:Yamada_feedback}) for $A=6.5$, $\kappa = 0.29$ and $\tau = 80$ (a), shows a stable non-lasing equilibrium (blue dot) and three attracting periodic orbits of different amplitudes (red, orange and green curves). Panel (b) shows the attractor map in the $(\Delta G , \Delta I )$-plane of small perturbations on gain $G$ and intensity $I$ from the non-lasing solution, were color corresponds to the corresponding attractor. White curves are third-order polynomials fitting the basins boundaries. Vertical black lines are the loci of three cross-sections for $\Delta G = -0.28$, $\Delta G = 0$ and $\Delta G = 0.2$, for which the basins boundaries are computed more precisely, as represented in panel (c), (d) and (e), respectively.}
\label{fig:PP and map basins tau 80}
\end{figure}

We focus here on the consequences of the observed high level of multistability. Specifically, we consider the example of phase portrait $\widetilde{15}$, in figure \ref{fig:PP and map basins tau 80}(a) for $(A,\kappa,\tau)=(6.5,0.29,80)$. Region $\widetilde{15}$ is adjacent to region 15 in the bifurcation diagram in figure \ref{fig:diag bif A6.5 large}, for $\kappa<0.3$, below the curve T of transcritical bifurcation. 
The stable solutions in this region include the equilibrium $E_1$ and three periodic orbits with different amplitudes and periods. As the three periodic orbits are very close to each other where their intensity $I$ is small, one can expect the system to be very sensitive to small perturbations. 
To investigate the features of multistable dynamics in more detail, we focus on the basins of attraction associated with the different attractors. In particular, knowing the basins' structure highlights low or high sensitivity of the system to the initial conditions.

In the experiment, when the laser is initially off, pulses are obtained by providing an external perturbation either on the gain $G$ or on the intensity $I$  \cite{Selmi_PRL_14}.
From a practical point of view, investigating the basins structure corresponds to determining the attractor on which the laser settles when it is perturbed from the off state. This question is addressed by time-domain simulations, in which we consider the equilibrium $E_1$ as initial condition and focus on the influence of small initial perturbations $\Delta G$ on the gain $G$ and $\Delta I$ on the intensity $I$, so that:
\begin{equation}
\begin{split}
&(G(t),Q(t),I(t))=(A,B,0),~~ \forall t \in [-\tau,0).\\
&(G(0),Q(0),I(0))=(A+\Delta G, B, \Delta I), ~ \text{ for } t=0.
\end{split}
\end{equation} 

For the parameters considered here, the system is close to the threshold $\kappa_T = 0.3$ at which $E_1$ becomes unstable, and a small but non-zero perturbation is enough to leave the basin of the off-state $E_1$.
Figure \ref{fig:PP and map basins tau 80}(b) represents in the $(\Delta G,\Delta I)$-plane the stable solution to which the system eventually settles in time-domain simulations, each solution being represented by the color of the corresponding attractor in the phase portrait in figure \ref{fig:PP and map basins tau 80}(a). As we deal here with a delay system, the basins associated with the different attractors live in infinitely many dimensions, and the attractor map in figure \ref{fig:PP and map basins tau 80}(b) is the trace of the basins structure in the $(\Delta G, \Delta I)$-plane.
The blue region is the basin of the equilibrium solution $E1$: if the intensity perturbation $\Delta I$ is smaller than a threshold which depends on the value of $\Delta G$, the laser relaxes back to its off-state without producing a pulse. The red region is the basin of the periodic orbit with the largest amplitude. Hence, a sufficiently large intensity perturbation $\Delta I$ always results in pulsations of the laser with a pulse repetition rate close to the delay time $\tau=80$. Again, the exact value of $\Delta I$ from which this basin is reached depends on the value of $\Delta G$. The same solution is observed for a sufficiently large, positive perturbation $\Delta G$, as long as $\Delta I$ is not too small. 
In between these two regions, the structure of the basins is much more intricate. The attractor map in figure \ref{fig:PP and map basins tau 80}(b) suggests that it is quite complicated in practice to predict both the transient and the long-term behaviour of the SLSA. 
The computation of the map in figure \ref{fig:PP and map basins tau 80} is particularly time-consuming. Not only is there a need for a fine mesh on both $\Delta G$ and $\Delta I$ to highlight the basins structure, but long transients are observed before the system settles to an attractor. In order to determine the attracting solution, simulations are thus run for a duration  $t_s = 250 \times \tau$. As a matter of fact, the computation of the map in figure \ref{fig:PP and map basins tau 80}(b) takes approximately a week with the method of steps for DDEs (implemented in C), on a desktop machine with three processors in parallel. 

Knowing the trace of the basins of attraction in the $(\Delta G,\Delta I)$-plane is not only interesting from an application point of view, but also from the perspective of the theoretical analysis of the model. Indeed, the complicated structure of the basin boundaries, highlighted in figure \ref{fig:PP and map basins tau 80}, gives insight into the structure of the stable manifolds of unstable solutions and, thus, a more global overview of the system dynamics. This is of particular interest because we deal here with a system with delay, with an infinite-dimensional phase space, meaning that it is not possible to compute the invariant manifolds with methods for ODEs \cite{krauskopf2005_IJBC,Krauskopf_Osinga_book_continuation}.

The attractor map in figure \ref{fig:PP and map basins tau 80}(b) clearly highlights the stripe structure of the basins, and suggests that it self-repeats at different scales when $\Delta I$ is varied.
To study the basins structure in more details, three cross-sections for $\Delta G = -0.28$, $\Delta G = 0$ and $\Delta G = 0.2$  are computed with a much finer mesh size of 0.0001 along $\Delta I$. 
These three sections, represented in panels (c), (d) and (e) of figure \ref{fig:PP and map basins tau 80}, clearly have a similar structure, which repeats at different scales for the three different values of $\Delta G$. Moreover, this highlights the self-repeating structure of the basins boundaries along $\Delta I$, with the same pattern repeating along the sections.
In panel (c) for example, the pattern found for $\Delta I \in [0.89,1.72]$, composed of a large red stripe and the upper green and orange stripes, repeats many times at smaller scales when $\Delta I $ decreases. The basins structure is thus particularly intricate for small $\Delta I$, where the pattern repeats at very small scales. 
In this pattern, the largest green stripe is bounded by two orange stripes, one at the top and the other at the bottom. Each of these orange stripes is in turn the lower and the upper limit, respectively, of other green stripes. Enlarging the cross-section in this region reveals that the same structure repeats again and again at increasingly smaller scale, leading to thiner and thiner stripes in very small region of the $(\Delta G, \Delta I)$-plane. This clearly shows that the boundaries of the basins associated with the three attracting periodic solutions accumulate on each other in a Cantor structure. It is interesting to note that this Cantor structure does not involve the basin of  the non-lasing equilibrium, which has a single boundary curve with the other basins in the $(\Delta G, \Delta I)$-plane. Note however, that the three basins boundaries of the periodic orbits accumulate on the basin boundary of the off-solution.
This self-repeating property along $\Delta I$, which is qualitatively independent of $\Delta G$, implies that the basins' boundaries are intermingled, which is a property that can be found in other complex systems without delay, including the Lorenz system \cite{Kennedy_93, Doedel_Lorentz_11}. The most obvious consequence is that the prediction of the asymptotic state of the laser may become practically impossible. Moreover, it also implies that trajectories can take a very long time before approaching a given attractor.

%
The three similar cross-sections in figure \ref{fig:PP and map basins tau 80} also allow us to determine the scale at which the pattern repeats along the whole range of $\Delta G$. In figure \ref{fig:PP and map basins tau 80}(b), the boundaries between the different stripes are fitted by third-order polynomials, given by:
\begin{equation}
\Delta I = \eta \left[ a+ b (\Delta G) + c (\Delta G)^2 +  (\Delta G)^3 \right].
\label{eq:polynomials}
\end{equation}
From the precise representation of the basins along the three cross-sections with fixed $\Delta G$, we estimate the set $(a,b,c)$ to obtain the best fit for each stripe boundary. We find that the stripe structure is represented by only two one-parameter families of third-order polynomials (\ref{eq:polynomials}), a family having fixed values of $a$, $b$ and $c$, and a single changing parameter $\eta$. More precisely, the family with $(a_1,b_1,c_1)=(-0.46, 2.03,-2.76)$ represents the uppermost stripes (\textit{i.e.}, the green and orange stripes with the largest values of $\Delta I$), and the family with $(a_2,b_2,c_2)=(0.92, -3.35, 2.5 )$ represents all of the other stripes. For each family, two polynomial boundary curves, corresponding to two different values of the coefficient $\eta$, are plotted in white in figure \ref{fig:PP and map basins tau 80}(b).

\begin{figure}[t!]
\begin{center}
\includegraphics[width=0.5\textwidth]{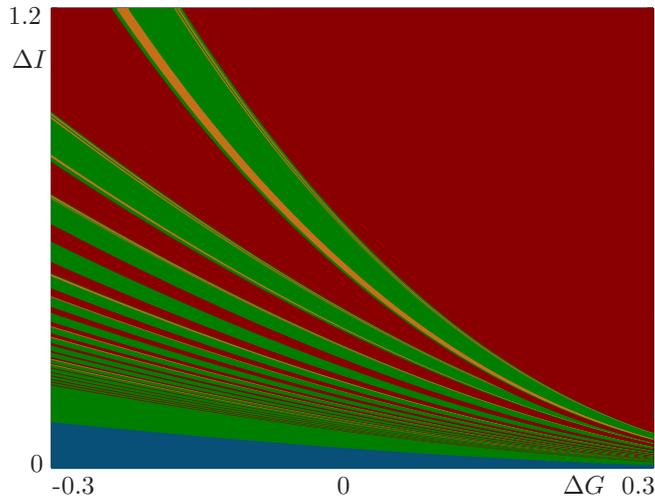}
\caption{A refined attractor map in the $(\Delta G , \Delta I )$-plane, obtained from the section $\Delta G =0$ in figure \ref{fig:PP and map basins tau 80}(d) and families of cubic polynomials given by equation (\ref{eq:polynomials}).}
\label{fig:rough and fitted map basins tau 80}
\end{center}
\end{figure}

The parameters $(a_1,b_1,c_1)$ and $(a_2,b_2,c_2)$ can then be use to extend the accurate cross-section $\Delta G=0$ to the whole range of $\Delta G$. In this way, we obtain a refined version of the attractor map without running CPU-consuming simulations on a finer mesh for the whole range of $\Delta G$ and $\Delta I$.
The resulting much finer map of the basins structure is shown in figure \ref{fig:rough and fitted map basins tau 80}. Clearly, the SLSA is less sensitive to perturbations for negative values of $\Delta G$. Conversely, for $\Delta G>0$ the range of $\Delta I$ in which the different basins are intricate is smaller; on the other hand in this range, the system is extremely sensitive to small perturbations and the behaviour of the SLSA thus easily becomes practically unpredictable.

\begin{figure}[t]
\includegraphics[width=\textwidth]{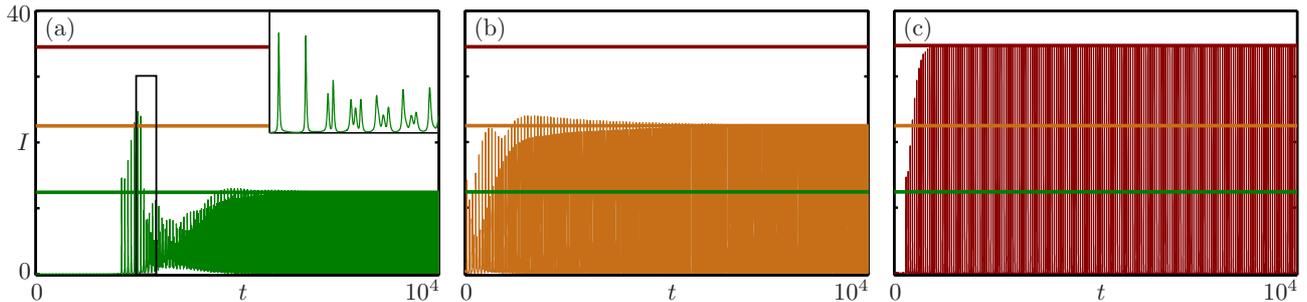}
\caption{Time series of the intensity $I$ of system (\ref{eq:Yamada_feedback}) for $(A,\kappa,\tau)=(6.5,0.29,80)$, $\Delta G = 0$, and for $\Delta I = 0.08$ (a), $\Delta I = 0.512$ (b) and $\Delta I = 0.2$ (c). The horizontal lines are the amplitudes of the three different stable periodic orbits of phase portrait \ref{fig:PP and map basins tau 80}(a). The inset in panel (a) is an enlargement of the transient in the black box. }
\label{fig:example transient map}
\end{figure}

Associated with the Cantor-set structure of the basins is a wealth of transient dynamics exhibited by the system. Figure \ref{fig:example transient map} presents three time series observed for $\Delta G = 0$ and $\Delta I = 0.08$, $\Delta I = 0.512$ and $\Delta I = 0.2$ in panels (a), (b) and (c), respectively. Each of the three values of $\Delta I$ leads to a given stable periodic solution of the phase portrait $\widetilde 15$ found in figure \ref{fig:PP and map basins tau 80}(a). The comparison between the three transients highlights the different time scales on which they take place. Moreover, as shown in the inset of figure \ref{fig:example transient map}(a), complicated dynamics can occur during the transients, including multipulses, where several pulses occur before the intensity drops back to zero.
Both the structure of the basins of attraction and the complex transient dynamics observed here in a deterministic configuration suggest that even a small amount of noise, as necessarily present in an experiment, may induce complicated, even practically unpredictable dynamics.
Moreover, in an actual experiment, the delay time can be up to $\tau =15000$ \cite{Felix_master_thesis}, and an even more complex configuration, with more stable and unstable periodic orbits, is to be expected \cite{Yanchuk}. One can then expect intermingled basins boundaries, involving many more basins, each of which is associated with one of the many attracting periodic orbits.

\section{Discussion\label{sec:discussion}}

We investigated the effect of delayed optical feedback on the dynamics of the Yamada model for self-pulsations in a semiconductor laser with saturable absorber (SLSA). 
The bifurcation analysis in the $(\tau,\kappa)$-plane of the delay time and feedback strength shows that the bifurcation diagram quickly becomes very complex when the delay $\tau$ is increased. The case of the SLSA in the non-excitable regime is particularly interesting: the zero-intensity equilibrium being the only solution when $\kappa =0$, one might expect that the feedback only has small consequences on the laser's dynamics. However, for nonzero feedback, the laser shows a wealth of new dynamics from intermediate values of $\tau$. 
This includes multistability of several pulse-like periodic solutions with different amplitudes and periods, as well as stable quasiperiodic oscillations on tori. In particular, the level of multistability increases significantly with the delay time $\tau$: more and more attracting self-pulsing solutions coexist, where repetition rates of the pulses are submultiples of $\tau$. 
We showed that these feedback-induced dynamics can be found in the $(A,\kappa)$-plane of the main control parameters, which means that they are accessible experimentally. 
These results show that, despite its simplicity, the Yamada model with feedback is able to produce a wealth of complex dynamics. 
Some of the scenarios observed in the Yamada model with feedback appear to correspond well to dynamics observed in an actual experiment, which at present time have been reported only in \cite{Felix_master_thesis,Terrien_OSA_QP}; an actual comparison between the model dynamics and experimental observations is the subject of on-going research.

We especially focused on features of multistability, and showed evidence of the Cantor-set like structure of the basins of the different pulsing solutions, by considering their trace in the $(\Delta G, \Delta I)$-plane of perturbations of gain $G$ and intensity $I$. It follows that the system is very sensitive to perturbations, especially when the intensity is small. A small perturbation can thus be sufficient to induce the system to jump between different basins. This suggests that for large values of the delay, as often considered in practice, the basins of the many coexisting stable periodic solutions might have intermingled boundaries. In particular, this means in practice that a small amount of noise may trigger unpredictable dynamics. 

High-sensitivity to noise might explain complex phenomena observed in an experiment, including multipulses and jumps between different stable solutions \cite{Felix_master_thesis}. An on-going study suggests that the varying life times of pulse trains observed in the experiment are explained by noise-induced transitions from a stable pulsing solution to the off-state equilibrium. The Yamada model with feedback and additional noise terms accurately reproduces this phenomenon, showing good agreement between the experimental and numerical escape times \cite{Barbay_pers_com}. In future work, we will focus on a more direct comparison between the experimental observations and the results from the theoretical analysis of the model. In particular, on-going experiments investigate some of the complex dynamics that we discussed above. 
It aims at showing experimental evidence of multistable dynamics, and associated characteristics of the attractors' basins. Although this last point is particularly challenging to demonstrate in an experiment, it can nevertheless be directly related to the observed sensitivity of the SLSA to noise and small perturbations. On-going experimental investigations also focus on quasiperiodic dynamics: because, in some cases, they may look similar to noisy periodically pulsing solution, such dynamics is quite difficult to identify experimentally \cite{Barbay_pers_com}.

Overall, our results suggest that the Yamada model accurately describes pulsing lasers with feedback as long as there is no other internal dynamics than pulsing or excitability. 
In future work, it would be interesting to compare the dynamics of the Yamada model with feedback with other types of pulsing lasers subject to feedback, in particular with passively mode-locked lasers. Although the underlying physical mechanism responsible for self-pulsations in passively mode locked lasers is different from Q-switching, which implies the use of different mathematical models, the dynamics of both kinds of lasers display interesting similarities when subject to feedback. As an example, the increasing level of multistability observed when the delay becomes larger was also highlighted for a passively mode-locked ring cavity laser with optical feedback: more and more pulsing solutions coexist, which correspond to different numbers of pulses in a cavity roundtrip \cite{Otto_NJP_2012, Jaurigue_14}. 
A recent study focused on multistable dynamics between several pulsing solutions and a non-lasing equilibrium in a passively mode locked laser subject to feedback \cite{marconi}. It showed that, when the delay becomes larger than the lasers internal timescales, pulses can be triggered  which almost behave as if they are independent from each other: arbitrary sequences of pulses can thus be addressed, where pulses are not necessarily regularly spaced. Such dynamics is often referred to, in the literature, as temporal cavity solitons, or localized structures.
Despite the model we consider here being quite different, a preliminary study suggests that both the Yamada model subject to feedback with large delay and an actual experiment display, phenomenologically, this type of dynamics \cite{Barbay_pers_com}. Moreover, some aspects of the dynamics of the Yamada model are very reminiscent of phenomena recently observed in neuromimetic excitable, slow-fast photonic systems subject to feedback \cite{Romeira_nature_16}. This type of device has also been shown to produce multiple localized structures. Overall, it suggests that the Yamada model might also be a minimal model suitable for the investigation of such phenomena.

An in-depth comparison between these different systems would be valuable to better understand the underlying mechanisms of such common dynamics, as well as to provide insight regarding the domain in which the Yamada model is valid. In particular, conversely to mode-locked lasers which may feature other intrinsic dynamics \cite{Jaurigue_15,Vladimirov_2005}, the device we consider here only displays excitability or pulsing in the absence of feedback. In this respect, the Yamada model is very similar to certain models of pulsing neurons \cite{Izhikevich_04,Selmi_PRL_14}.

The extended bifurcation analysis of a pulsing SLSA with self-feedback presented here might, thus, be of wider interest to the applied dynamical systems community. Especially, it can be seen as a first step in the investigation of more general delay-coupled pulsing systems. The next step would be the case of two delay-coupled pulsing SLSA, where the system studied here appear as a special case where the two lasers are identical. While coupled lasers have been extensively studied both experimentally and theoretically, it would be particularly interesting to focus on the effect of delayed coupling when the lasers are operating strictly in the pulsing regime. Again, such a configuration is closely related to that of connected excitable neurons.

\subsection*{Acknowledgement}
The authors thank Sylvain Barbay and Kathy Luedge for stimulating discussions, as well as the referees for helpful comments and suggestions. The research of S. T. has been funded by a Postdoctoral Fellowship of The Dodd-Walls Center for Photonic and Quantum Technologies.

\clearpage
\newpage

\appendix
\section{Parameters value}
\label{sec:appendix}
\begin{center}
Fixed parameters:  $B=5.8$ ; $\gamma = 0.04$ ; $a=1.8$.
 \vspace{5mm}
 
\begin{tabular}{|c|c|c|c|c||c|c|c|c|c|}
\hline
figure & A & $\kappa$ & $\tau$ &  figure & A & $\kappa$ & $\tau$  \\
\hline
\hline
2 & 6.5 & ~ & ~ & 5 panel 18 & 6.5 & 0.38 & 80.5  \\ 
\hline
3 panel 3 & 6.5 & 0.16 & 2.6 & 5 panel 19 & 6.5 & 0.38 & 90.5 \\
\hline
3 panel 5 & 6.5 & 0.25 & 38 & 6 panel (a) & 6.5 & 0.38 & ~ \\ 
\hline
3 panel 6 & 6.5 & 0.25 & 25 & 6 panels (b), (c) and (d) & 6.5 & 0.38 & 100  \\
\hline
3 panel 7 & 6.5 & 0.38 & 21  &7 panels (a) & 6.5 & 0.38 & 117  \\
\hline
3 panel 8 & 6.5 & 0.38 & 33 & 7 panels (b) & 6.5 & 0.38 & 140 \\
\hline
3 panel 10 & 6.5 & 0.25 & 52  & 8 & 5.9 & ~ & ~  \\
\hline
3 panel 11 & 6.5 & 0.38 & 43 & 9 panel (a) & 5.9 & 0.56 & 102  \\
\hline
3 panel 12 & 6.5 & 0.38 & 55  & 9 panel (b) & 5.9 & 0.55 & 117  \\
\hline
3 panel 13 & 6.5 & 0.0152 & 26.8 &10 panel (a) & ~ & ~ & 25  \\
\hline
4 & 6.5 & ~ & ~ & 10 panel (b) & ~ & ~ & 40 \\
\hline
5 panel 14 & 6.5 & 0.38 & 63  &  10 panel (c) & ~ & ~ & 58   \\
\hline
5 panel 15 & 6.5 & 0.38 & 68   & 11 & 6.5 & 0.29 & 80 \\
\hline
5 panel 16 & 6.5 & 0.38 & 75   & 12 & 6.5 & 0.29 & 80\\
\hline
5 panel 17 & 6.5 & 0.38 & 79   & 13 & 6.5 & 0.29 & 80 \\
\hline

\end{tabular}

\end{center}

\newpage
\bibliographystyle{plain}
\bibliography{bib_SLSA}

\end{document}